\documentclass[journal]{IEEEtran}
\IEEEoverridecommandlockouts
\usepackage{amsfonts}
\usepackage{amssymb}
\usepackage{amsmath}
\usepackage{diagbox}
\usepackage[ruled]{algorithm2e}
\usepackage{makecell} 
\usepackage{multirow}
\usepackage{xcolor}
\usepackage{graphicx}
\usepackage{cite}
\usepackage{exscale}
\usepackage{relsize}
\usepackage{algpseudocode}
\usepackage{graphics}
\usepackage{mathrsfs}
\usepackage{siunitx}
\usepackage{threeparttable}
\usepackage{url}
\usepackage{booktabs}
\usepackage{lipsum}
\usepackage{array}
\usepackage{comment}
\DeclareMathOperator*{\argmax}{argmax}

\newcommand{\bm}[1]{\mbox{\boldmath{$#1$}}}

\begin{document}
\title{Integrated Sensing, Communication, and Over-The-Air Control of UAV Swarm Dynamics

\author{Zhuangkun Wei, Wenxiu Hu, Yathreb Bouazizi, Mengbang Zou, Chenguang Liu, \\Yunfei Chen, Hongjian Sun, Julie McCann

\thanks{Zhuangkun Wei, Chenguang Liu, Yunfei Chen, and Hongjian Sun are with the Department of Engineering, Durham University, DH1 3LE, UK. (Email: zhuangkun.wei/chenguang.liu/yunfei.chen/hongjian.sun@durham.ac.uk) \\
Yathreb Bouazizi and Julie McCann are with the Department of Computing, Imperial College London, SW7 2RH, UK. (Email:y.bouazizi18/j.mccann@imperial.ac.uk)\\
Mengbang Zou is with the School of Aerospace, Transport, and Manufacturing, Cranfield University, MK43 0AL, UK. (Email: m.zou@cranfield.ac.uk)\\
Wenxiu Hu is with the Optoelectronics Research Centre, Southampton University, SO17 1BJ, UK. (Email: w.hu@soton.ac.uk) \emph{Corresponding Author: Wenxiu Hu}\\
This work is supported by the Engineering and Physical Sciences Research Council: Communications Hub For Empowering Distributed ClouD Computing Applications And Research (CHEDDAR) with grant id: EP/X040518/1 and EP/Y037421/1.}}
}

\maketitle 

\begin{abstract}
Coordinated controlling a large UAV swarm requires significant spectrum resources due to the need for bandwidth allocation per UAV, posing a challenge in resource-limited environments. Over-the-air (OTA) control has emerged as a spectrum-efficient approach, leveraging electromagnetic superposition to form control signals at a base station (BS). However, existing OTA controllers lack sufficient optimization variables to meet UAV swarm control objectives and fail to integrate control with other BS functions like sensing. This work proposes an integrated sensing and OTA control framework (ISAC-OTA) for UAV swarm. The BS performs OTA signal construction (uplink) and dispatch (downlink) while simultaneously sensing objects. Two uplink post-processing methods are developed: a control-centric approach generating closed-form control signals via a feedback-looped OTA control problem, and a sensing-centric method mitigating transmission-induced interference for accurate object sensing. For the downlink, a non-convex problem is formulated and solved to minimize control signal dispatch (transmission) error while maintaining a minimum sensing signal-to-noise ratio (SNR). Simulation results show that the proposed ISAC-OTA controller achieves control performance comparable to the benchmark optimal control algorithm while maintaining high sensing accuracy, despite OTA transmission interference. Moreover, it eliminates the need for per-UAV bandwidth allocation, showcasing a spectrum-efficient method for cooperative control in future wireless systems.
\end{abstract}


\section{Introduction}
The emergent Internet of Everything (IoE) and intelligent edge learning systems will encompass vast numbers of devices that require management and control for critical applications in a coordinated manner, such as smart-grid synchronization \cite{8166737}, drone formation flight \cite{10444964}, and platoon driving \cite{jia2016platoon}, etc. The limited availability of spectrum resources, however, presents a challenge in allocating bandwidth to a large number of networked devices for coordinated management. This issue becomes particularly severe in cooperative control frameworks \cite{8166737, lunze2014control} and multi-UAV trajectory optimization scenarios \cite{9678115}, where each terminal requires dedicated frequency bandwidth for information exchange including reporting its state to a coordinate controller and to receive control signals. Actually, due to this limited spectrum, current UAV swarm displays are restricted to predefined trajectories for each UAV, rather than allowing real-time path determination, not to mention UAV coordination or networking for more complex tasks.

\begin{table}[!t]
\centering
\caption{Conceptual Comparison with other ISAC, OTA, and UAV works}
\label{table0}
\begin{tabular}{|p{2.4cm}|p{1.45cm}|p{2cm}|p{1.3cm}|}
\hline
{\bf Research Categories} & {\bf Number of required channels} & {\bf Feedback-looped control objectives} & {\bf Sensing in the same frequency} \\[1mm]
\hline
UAV trajectory optimization & Number of UAVs & \centering{$\times$} & {~~~~~$\times$}\\
\hline
ISAC-UAV trajectory optimization & Number of UAVs & \centering{$\times$} & ~~~~~\checkmark \\
\hline
OTA-FL & \centering{1} & \centering{$\times$} & {~~~~~$\times$} \\
\hline
ISAC OTA-FL & \centering{1} & \centering{$\times$} & ~~~~~\checkmark \\
\hline
Existing control theory: LQR, MPC, etc & Number of UAVs & \centering{\checkmark} & {~~~~~$\times$}\\
\hline
{OTA-Control.} & \centering{1} & Current versions cannot work for UAV swarm & {~~~~~$\times$} \\
\hline
ISAC-OTA Control. (proposd) & \centering{1} & \centering{\checkmark} & {~~~~~\checkmark} \\
\hline
\end{tabular}
\end{table}

To address the spectrum scarcity challenges, the concept of over-the-air (OTA) computation has been proposed. OTA computations allow multiple devices to use the same time and frequency block to transmit their signals to a server or base-station (BS), where the signal processing of transmitted signals can be directly performed by tuning the electromagnetic waves in the air. In doing so, OTA computation can save the spectrum resources for a wide range of linear operations in IoE and edge computing system. One popular example lies in the designs of OTA federated learning (OTA-FL) \cite{Yang2020OTAFL}, where the linear aggregation of neural network (NN) parameters transmitted by selected terminal clients are pursued leveraging the linear superposition of appropriately tuned wireless channels. Several techniques have been developed to minimize the mean squared error (MSE) between channel coefficients and linear FL aggregation
weights (e.g., FedAvg). These include but are not limited to transmitter precoding and beamforming designs \cite{sery2021over}, BS or server beamforming \cite{kim2023beamforming}, movement and trajectory optimization of mobile terminals \cite{10283588}, reconfigurable intelligent surface (RIS) phase adjustment \cite{10649032,zheng2022balancing}, and recent advanced fluid antennas \cite{ahmadzadeh2024enhancement}.
However, the established OTA-FL schemes cannot be extended to multi-device cooperative control systems \cite{9641840}. This is because OTA-FL are not designed to address the complex feedback-looped objective function of control signals, which involves the accumulated effects of future predicted system states based on the feedback from the current control signals. Moreover, the feedback-looped objective function creates a control layer state dependency among different terminals, which differs from the independence or spatiotemporal correlations typically assumed in FL.

To exploit the OTA concept in control theory, OTA controllers have been proposed in \cite{9500236,9641840}. Compared with traditional control frameworks, e.g., linear quadratic regulator (LQR), and model predictive control (MPC), which require the allocation of bandwidth for different terminals, OTA controller allows all network nodes to transmit their observed states on the same time and frequency block. Then, the received signals at the BS (serving as the coordinate controller) form the control signals through the linear superposition of compensated wireless channels. In \cite{9641840}, a terminal transmitter (Tx) power-allocation based OTA controller was proposed, aiming to find the optimal transmission powers of network terminals to minimize the $H_{\infty}$ based OTA control objective. However, as the network state-space and control signal space scale up, the size of the OTA control operator, defined by the product of these dimensions, far exceeds the number of Tx-power optimization variables provided by the antenna-limited terminals. For instance, consider a cooperative control scenario of $50$ UAVs, each of which has $6$ states i.e., xyz-positions and velocities, and $3$ control variables i.e., xyz-accelerations. The terminal Tx-power-based OTA controller can only have $6^2\times50=1800$ precoding variables, insufficient to configure the linear control operator of size $(3\times50)\times(6\times50)=45,000$ that maps a total of $6\times50$ states to $3\times50$ control signals, leading to significant control errors and slow convergence (shown later in Fig. \ref{figs1}).

Besides, current OTA controller designs do not integrate the coordinate controller role with the diverse multi-functionalities of BS for future wireless systems.
The BS is envisioned to have diverse functionalities, such as communications, computations and sensing. Thus, the integrated sensing and communication (ISAC) concept has emerged as a promising enabler for future 6G and beyond wireless networks \cite{ISAC_6G_Survey1, ISAC_6G_Survey2, ISAC_App_Survey}. ISAC enables the cohabitation of communication and sensing functionalities in the same hardware system. This integration leverages orthogonal subspace division (loose coupling) \cite{ISAC_InformationTheory1}, ranging from time frequency and code \cite{Time_Division_ISAC_Transport, Access_Mangement_DJSC, ISAC_Waveform_Design, Code_Division_ISAC}, to spatial beamforming designs \cite{ISAC_RessourceAlloc,10135096,ISAC_SignalProcessing}. Using different levels of integration, ISAC offers a great potential to shape the future of several applications including smart healthcare \cite{ISAC_Health}, smart transportation \cite{ISAC_Transport_Vehicle,ISAC_Transport_Drones}, smart industry and critical infrastructure monitoring \cite{Smart_Indust}. 

The performance gains achieved by each of the ISAC and the OTA computations make their integration appealing. The instances include the framework designs of the over-the-air integrated sensing communication and computation (Air-ISCC) \cite{OTA_ISCAC1}, the integrated sensing and OTA computation (ISAA) \cite{Integrated_Sensing_OTAC}, and ISAC for OTA-FL (Fed-ISCC) \cite{10621049}. Motivated by these designs, a novel framework to integrate sensing and OTA controller for spectrum efficient networked dynamics control is necessary for large-scale IoE control scenarios, such as UAV swarm and smart-grid.

In this work, we propose a novel integrated sensing and OTA control framework, namely ISAC-OTA controller, for UAV swarm system. Compared with other related works in Table. \ref{table0}, the design of ISAC-OTA controller is able to realize object sensing and coordinate feedback-looped control of dynamical UAV swarm system in the same frequency. The main contributions are as follows:

(1) We formulate an ISAC-OTA controller framework, where BS serves as the coordinate controller of the dynamical UAV swarm system, as well as performing the object sensing tasks, within the same time and frequency block. In the UAV-BS uplink, BS is designed to (i) construct the overall control signal for UAV swarm from the OTA transmitted UAV states corrupted by the echoed radar signal interference, and (ii) perform the sensing task with uplink transmission interference. In the BS-UAV downlink process, BS is designed to dispatch its OTA-constructed control signals dedicated to each UAV, and maintain the sensing performance that leverages both the downlink transmission signal and radar sensing signal. 

(2) In the uplink process, we design two post-processing matrices at BS: namely control-centric and sensing-centric matrices, to avoid the interference between OTA state transmission and echoed radar signal. For UAV coordinate control purposes, a feedback-lopped control objective function is formulated, incorporating the wireless channels and the control-centric BS post-processing matrix into the accumulated state error minimization problem. Then, we convert the complex feedback-looped objective function to a mini-max eigenvalue problem, and deduce the optimal closed-form control signal. For sensing purposes, to mitigate the interference from uplink state transmission, the sensing-centric post-processing matrix is derived via the orthogonal subspace to the uplink transmission channels. 

(3) In downlink designs, an optimization problem is formulated to minimize the transmission error of control signals, as well as maintain the minimum BS sensing signal-to-noise ratio (SNR). Unlike the download process in OTA-FL (e.g., Fed-ISCC \cite{10621049}), where every client receives the same data (i.e., global NN parameters), each UAV must be assigned only its portion of the overall control signals. This requires more complex designs and optimization for the BS precoding dispatch matrix and the BS sensing beamforming vector. To address the non-convex issue in the formulated downlink optimization problem, a successive convex approximation (SCA) based iterative algorithm is designed to find numerical solutions of optimal (sub-optimal) BS precoding matrix and sensing beamforming vector. 

(4) We evaluate our proposed ISAC-OTA controller in the coordinate control scenario with $50$ UAVs. The results show that the ISAC-OTA controller can (i) achieve comparable control performance (in terms of error and time convergence) to the benchmark LQR algorithm, without the need for bandwidth allocation to each UAV, and (ii) offer remarkable sensing accuracy despite OTA transmission interference.

The rest of this work is structured as follows. The networked dynamical system model of UAV swarm, and the wireless communication model are described in Section \ref{model}. The overall schematics of our ISAC-OTA controller framework is elaborated in Section \ref{schematic}. In Sections \ref{optimization}-\ref{downlink}, the uplink and downlink optimizations for ISAC-OTA controller are detailed. The simulation results are shown and analyzed in Section \ref{sim}. We finally conclude our work in Section \ref{conclusion}.

In the following, we use bold lower-case letters for vectors, and bold capital letters for matrices. We use $\|\cdot\|_2$ to denote the $2$-norm, $\|\cdot\|_F$ to denote the Frobenius norm, and $\text{diag}(\cdot)$ to diagonalize a vector. $|\cdot|$ represents the absolute value of a complex value. The matrix transpose, conjugate transpose, element-wise conjugate, Kronecker product, trace, and vectorize operators are denoted as $(\cdot)^T$, $(\cdot)^H$, $(\cdot)^*$, $\otimes$, $\text{tr}(\cdot)$, and $\text{vec}(\cdot)$. $\lambda_\text{max}(\cdot)$ represents the largest eigenvalue of a matrix. $\mathbf{I}_N$ represents the identity matrix with size $N\times N$. $\mathbb{E}(\cdot)$ represents the expectation. $\Re\{\cdot\}$ denotes to take the real part of a complex number. $\mathcal{CN}(\mu,\sigma^2)$ is to represent the complex Gaussian distribution with mean $\mu$ and variance $\sigma^2$.

\section{System Model}\label{model}
\begin{figure*}[!t]
\centering
\includegraphics[width=6.8in]{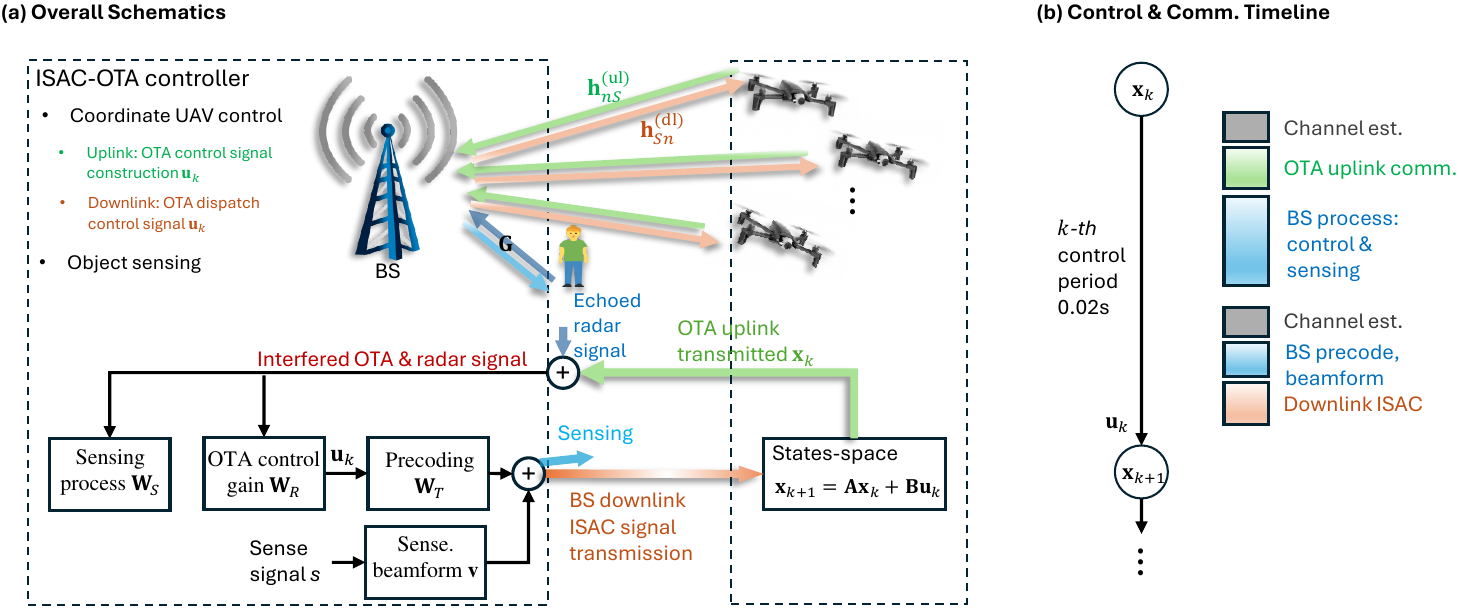}
\caption{Schematic flow of ISAC-OTA controller.
}
\label{figm1}
\end{figure*}

In this work,  we consider a cooperative control scenario of $N\in\mathbb{N}^+$ UAVs, where a BS serves as the coordinate controller for the UAV swarm dynamic system, as well as for the object sensing, as is shown in Fig. \ref{figm1}(a).

Traditionally, for coordinated control purposes, each UAV is typically allocated a unique bandwidth as an orthogonal channel for state transmission and control signal reception. To save bandwidth resources, we leverage the concept of an over-the-air controller \cite{9641840}. The OTA-controller allows all nodes to transmit their observed state values to the coordinate controller, through the same frequency over the air (without any digital modulation, and time or frequency division). These state values are sent via electromagnetic waves that are superposed with different channel coefficients, which, if properly tuned, allow the direct obtaining of the cooperative control signals at the BS. Here, we design a novel ISAC-OTA controller, which leverages the strong channel compensation ability of the BS, to integrate OTA control and sensing in the same time and frequency block to control the UAV swarm.

\subsection{Control Model of UAV Swarm Dynamical System}
The networked dynamical system is the coordinated control scenario of $N$ UAVs. Each $n$-th UAV has $6$ states (xyz-positions and velocities), denoted as $\mathbf{x}_k^{(n)}\in\mathbb{R}^6$, and are controlled by the motor force induced xyz-accelerations, denoted as $\mathbf{u}_k^{(n)}\in\mathbb{R}^3$, where $k\in\mathbb{N}^+$ is the controlling time step. The state-space UAV dynamic model is \cite{tahir2019state}:
\begin{equation}
\begin{aligned}
\mathbf{x}_{k+1}^{(n)}&=
\underbrace{\begin{bmatrix}
\mathbf{I}_3 & \Delta_t\mathbf{I}_3\\
\mathbf{O}_{3\times3} & \mathbf{I}_3
\end{bmatrix}}_{\overline{\mathbf{A}}}
\mathbf{x}_{k}^{(n)}+\underbrace{
\begin{bmatrix}
\Delta_t^2/2\mathbf{I}_3\\
\Delta_t\mathbf{I}_3
\end{bmatrix}
}_{\overline{\mathbf{B}}}\mathbf{u}_k^{(n)},\\
\mathbf{y}_{k+1}^{(n)}&=\overline{\mathbf{C}}\cdot\mathbf{x}_{k+1}^{(n)}+\mathbf{n}_{k+1}^{(n)}
\end{aligned}
\end{equation}
where $\Delta_t$ is the control interval. Here, the UAV dynamical system contains only their xyz positions to accelerations that will be coordinated with each other; the conversion of obtained xyz-acceleration to roll, pitch, and yaw rates and motor forces are locally pursued in each UAV \cite{luukkonen2011modelling}. $\mathbf{y}_{k+1}^{(n)}$ represents the observation of $n$-th UAV to measure its state at $(k+1)$-th controlling period, with the observation matrix $\mathbf{C}=\mathbf{I}_6$ and the observation noise $\mathbf{n}_{k+1}^{(n)}\sim\mathcal{N}(0,\sigma_n^2)$. 

To coordinately control $N$ UAVs, we specify the overall networked dynamical system by stacking states, control signals, observed states, and noises of each UAV, i.e., $\mathbf{x}_k=[(\mathbf{x}_k^{(1)})^T,\cdots,(\mathbf{x}_k^{(N)})^T]^T$, $\mathbf{u}_k=[(\mathbf{u}_k^{(1)})^T,\cdots,(\mathbf{u}_k^{(N)})^T]^T$, $\mathbf{y}_k=[(\mathbf{y}_k^{(1)})^T,\cdots,(\mathbf{y}_k^{(N)})^T]^T$, and $\mathbf{n}_k=[(\mathbf{n}_k^{(1)})^T,\cdots,(\mathbf{n}_k^{(N)})^T]^T$. The overall networked dynamical state-space model can be then expressed as: 
\begin{equation}
\label{ode1}
\begin{aligned}
\mathbf{x}_{k+1}&=\mathbf{A}\cdot\mathbf{x}_k+\mathbf{B}\cdot\mathbf{u}_k,\\
\mathbf{y}_{k+1}&=\mathbf{C}\cdot\mathbf{x}_{k+1}+\mathbf{n}_{k+1}
\end{aligned}
\end{equation}
where $\mathbf{A}=\mathbf{I}_{N}\otimes\overline{\mathbf{A}}$, $\mathbf{B}=\mathbf{I}_{N}\otimes\overline{\mathbf{B}}$, and $\mathbf{C}=\mathbf{I}_N\otimes\overline{\mathbf{C}}$.

\subsection{Wireless Channel Model}
The implementation of both OTA control of networked dynamical system and object sensing requires the uplink and downlink processes. 
Each UAV is equipped with a single antenna to transmit its observed state to the BS in the uplink, and receive the control signal from the BS in the downlink. The BS is equipped with a half-wavelength spaced uniform linear array (ULA), which consists of $N_r\in\mathbb{N}^+$ Rx antennas to receive uplink OTA constructed control signals and radar echo signals for sensing, and $N_t\in\mathbb{N}^+$ Tx antennas to send downlink integrated control and sensing signals. 

The uplink channel from the $n$-th UAV to BS, denoted as $\mathbf{h}_{nS}^\text{(ul)}\in\mathbb{C}^{N_r\times1}$, and the downlink channel from BS to the $n$-th UAV, denoted as $\mathbf{h}_{Sn}^\text{(dl)}\in\mathbb{C}^{1\times N_t}$, are formulated via the narrow-band geometric models, i.e., \cite{9300189}
\begin{equation}
\label{eq02}
\begin{aligned}
\mathbf{h}_{nS}^\text{(ul)}=\rho_{nS,1}\cdot\bm{\psi}(\theta_{nS,1},N_r)+\sum_{l=2}^L\rho_{nS,l}\cdot&\bm{\psi}\left(\theta_{nS,l},N_r\right),\\
\mathbf{h}_{Sn}^\text{(dl)}=\rho_{Sn,1}\cdot\bm{\psi}^T(\theta_{Sn,1},N_t)+\sum_{l=2}^L\rho_{Sn,l}\cdot&\bm{\psi}^T\left(\theta_{nS,l},N_t\right)
\end{aligned}    
\end{equation}
In Eq. (\ref{eq02}), $L$ is the number of paths including $1$ line-of-sight (LoS) and $L-1$ non-LoS (NLoS) paths \cite{9300189}. $\rho_{nS,1}=\rho_{Sn,1}=\sqrt{C_0\cdot d_{nS}^{-\eta_\text{LoS}}}$ are the gains of the LoS paths between UAV $n$ and BS, and $\rho_{nS,l},\rho_{Sn,l}\sim\mathcal{CN}\left(0,C_0\cdot d_{nS}^{-\eta_\text{NLoS}}/(L-1)\right),l=2,\cdots,L$ are the $l$-th NLoS path gains from UAV $n$ to BS and from BS to UAV $n$, respectively, with $d_{nS}$ the distance between UAV $n$ and BS, $C_0$ the path loss at the reference distance of $1$m, and $\eta_\text{LoS}$ and $\eta_\text{NLoS}$ the exponential LoS and NLoS loss factors. $\theta_{nS,l},\theta_{Sn,l}\in[-\pi/2,\pi/2]$ are the half-space azimuth angles of the $l$-th path arriving at (uplink) and departing from (downlink) BS.
$\bm{\psi}(\theta,M)\triangleq[1, \exp(j\pi\cos\theta), \cdots,\exp(j\pi M\cos\theta)]^T$ defines the spatial channel vector of BS ULA.

For the BS sensing task, we consider the echoed radar response model to characterize the radar response matrix of a target \cite{8579200}, i.e., 
\begin{equation}
\mathbf{G}=\beta\cdot\bm{\psi}(\alpha_o, N_r)\cdot\bm{\psi}^T(\alpha_o, N_t)
\end{equation}
where $\beta$ is the complex channel gain that includes both the round-trip path-loss and the radar cross section (RCS), and $\alpha_o$ is the angle of the sensing object that will be estimated jointly in this work with the OTA control objective.

\section{Schematic of ISAC-OTA Controller}\label{schematic}
We consider a total of $K\in\mathbb{N}^+$ controlling periods. In each controlling period, BS will (i) generate control signals from the received UAV states in the uplink, (ii) dispatch the generated control signal to each UAV, and (iii) perform the object sensing task.

\subsection{UAV-BS Uplink}
The UAV-BS uplink aims to transmit all UAV states through OTA mode in the same frequency, to BS to generate the overall control signal $\mathbf{u}_k$. Before the $k$-th controlling period, uplink channel estimation is conducted at BS when it receives the pilot sequences transmitted by all UAVs in the same frequency. 

At the start of the $k$-th controlling period, all UAVs transmit their observed $6$-state in $6$ time division to the BS, i.e., $\mathbf{y}_k^{(n)}=\mathbf{x}_k^{(n)}+\mathbf{n}_k^{(n)}$ with $\mathbf{x}_k^{(n)}=[x_{k,1}^{(n)},x_{k,2}^{(n)},x_{k,3}^{(n)},x_{k,4}^{(n)},x_{k,5}^{(n)},x_{k,6}^{(n)}]^T$. BS then receives $\mathbf{r}_k^{\text{(ul)}}[1]$, $\mathbf{r}_k^{\text{(ul)}}[2]$, $\mathbf{r}_k^{\text{(ul)}}[3]$, $\mathbf{r}_k^{\text{(ul)}}[4]$, $\mathbf{r}_k^{\text{(ul)}}[5]$, and $\mathbf{r}_k^{\text{(ul)}}[6]$ of $6$ successive uplink transmission time-slots, i.e., 
\begin{equation}
\label{receive1}
\begin{aligned}
\mathbf{r}_k^{\text{(ul)}}[t]
=\sum_{n=1}^{N}\mathbf{h}_{nS}^{\text{(ul)}}\cdot\gamma^\text{(ul)}\cdot x_{k,t}^{(n)}+&\mathbf{G}\cdot s[t]+\bm{\varepsilon}_k^{\text{(ul)}}[t],\\
&t=1,\cdots,6,
\end{aligned}
\end{equation}
where $\gamma^\text{(ul)}$ is the predefined uplink scaling factor to constrain transmission power, i.e.,  $|\gamma^\text{(ul)}y_{k,t}^\text{(n)}|\leq1$W, and is known to all UAVs and BS. 
In Eq. (\ref{receive1}), $\mathbf{G}\cdot s[t]$ represents the received echoed radar signal $s[t]\in\mathcal{CN}(0,1)$ at $t$ time-slot. The existence of the echoed radar signal in the ISAC uplink is due to the radar's continuous operation, as it persistently transmits signals for sensing purposes.
$\bm{\varepsilon}_k^\text{(ul)}[t]\sim\mathcal{CN}(0,\sigma^2\mathbf{I}_{N_r})$ is the noise vector that involves received noise and observation noise, with variance $\sigma^2$. 

After the BS receives all $6$ signals, the BS stacks all of them as $\mathbf{r}_k^{\text{(ul)}}$, i.e., 
\begin{equation}
\label{receive}
\mathbf{r}_k^{\text{(ul)}}=\text{vec}\left(\left[\mathbf{r}_k^{\text{(ul)}}[1], \mathbf{r}_k^{\text{(ul)}}[2], \mathbf{r}_k^{\text{(ul)}}[3], \mathbf{r}_k^{\text{(ul)}}[4], \mathbf{r}_k^{\text{(ul)}}[5], \mathbf{r}_k^{\text{(ul)}}[6]\right]\right).
\end{equation}
The received signal $\mathbf{r}_k^\text{(ul)}[t],~t=1,\cdots,6$ and its stacked version $\mathbf{r}_k^{\text{(ul)}}$, will be used to (i) construct the OTA control signal $\mathbf{u}_k$, and (ii) perform object sensing task. The challenge lies in that the uplink state transmission and radar sensing signal interfere with each other. The concept of OTA signal processing applies here, whereby two BS post-processing matrices $\mathbf{W}_R$ and $\mathbf{W}_S$ will be designed in Section \ref{optimization}: $\mathbf{W}_R$ is to compensate the uplink channel coefficients to directly construct control signal $\mathbf{u}_k$, and $\mathbf{W}_S$ is to derive an accurate sensing signal by projecting to the orthogonal transmission subspace. The post-processed results, i.e.,  
\begin{subequations}
\label{ul}
\begin{align}
&\mathbf{u}_k=\mathbf{W}_R\cdot\mathbf{r}_k^\text{(ul)},\label{ul-u}\\
&\mathbf{q}[t]=\mathbf{W}_S\cdot \mathbf{r}_k^\text{(ul)}[t],~t=1,\cdots,6\label{ul-s}
\end{align}
\end{subequations}
are the constructed OTA control signal, and the input of the MUSIC algorithm for object sensing, respectively.

\subsection{BS-UAV Downlink Procedure}\label{BS-UAV}

After BS constructs the OTA control signal $\mathbf{u}_k$ for $k$-th controlling period, BS will (i) send UAVs' their own control signal via the OTA mode using the same frequency, as well as (ii) maintain the object sensing task. %Before the downlink procedure, downlink channel estimation is pursued at BS via standard channel estimation process. 

Recall that $\mathbf{u}_k=[(\mathbf{u}_k^{(1)})^T,\cdots,(\mathbf{u}_k^{(N)})^T]^T$, and each $\mathbf{u}_k^{(n)}=[u_{k,1}^{(n)},u_{k,2}^{(n)},u_{k,3}^{(n)}]^T$ contains $3$ signals. BS therefore will split and send the whole $\mathbf{u}_k$ into $3$ parts in $3$ successive downlink transmission time-slots, denoted as $\tilde{\mathbf{u}}_k[t]\triangleq[(u_{k,t}^{(1)})^T,(u_{k,t}^{(2)})^T,\cdots,
(u_{k,t}^{(N)})^T]^T,~t=1,2,3$. Then, as is integrated with the radar sensing purposes, The BS downlink ISAC signal is expressed as:
\begin{equation}
\label{isac_signal1}
\bm{\zeta}_k[t]=\mathbf{W}_T\cdot\tilde{\mathbf{u}}_k[t]
+\mathbf{v}\cdot s[t],~t=1,2,3,
\end{equation}
where $\mathbf{W}_T$ is the BS precoding matrix, and $\mathbf{v}$ is the 
sensing beamforming vector. The designs of $\mathbf{W}_T$ and $\mathbf{v}$ are required to balance both the control signal transmission accuracy at each UAV, and the radar sensing performance at BS, which will be detailed in Section \ref{downlink}.

The received signals at $n$-th UAV of the corresponding $3$ successive transmission time-slots are denoted as $z_k^{(n)}[1]$ $z_k^{(n)}[2]$ and $z_k^{(n)}[3]$, which are expressed as:
\begin{equation}
z_k^{(n)}[t]=\mathbf{h}_{Sn}^\text{(dl)}\cdot\bm{\zeta}_k[t]+\epsilon_k^{(n)}[t],~t=1,2,3,
\end{equation}
where $\epsilon_k^{(n)}[t]\sim\mathcal{CN}(0,\sigma^2)$ is the received noise. The $n$-th UAV then constructs its control signal $\mathbf{u}_k^{(n)}$ by scaling its received signals via the scaling factor $a$, i.e., 
\begin{equation}
\label{uav_receive1}
\mathbf{u}_k^{(n)}=a\cdot \left[z_k^{(n)}[1],z_k^{(n)}[2],z_k^{(n)}[3]\right]^T.
\end{equation}
Here, the scaling factor is optimized in BS and sent in advance to all UAVs in an error-free manner \cite{9791337}. 
The constructed control signal $\mathbf{u}_k^{(n)}$ will be further input to the local controller of UAV $n$ to compute the row pitch and yaw rates, and the forces of $4$ rotors \cite{luukkonen2011modelling}. 

For sensing at BS, its received signal is from the continuous radar signal echoed by the object of the interest, i.e., 
\begin{equation}
\label{receive33}
\mathbf{r}_k^{\text{(dl)}}[t]=\mathbf{G}\cdot\bm{\zeta}_k[t]+\bm{\varepsilon}_k^\text{(dl)}[t],~t=1,2,3,
\end{equation}
where $\bm{\varepsilon}_k^\text{(dl)}[t]\sim\mathcal{CN}(0,\sigma^2\mathbf{I}_{N_r})$ is BS received noise vector. MUSIC algorithm (detailed in Appendix \ref{Appendix1}) will then be employed for sensing tasks.

\section{Uplink ISAC-OTA Control Designs}\label{optimization}
In this section, we elaborate on the designs of optimizations for the uplink OTA control signal construction, as well as maintaining the integrated sensing operation. From Eq. (\ref{ul}), the optimization in the UAV-BS uplink process contains the BS post-processing matrices, i.e., (i) $\mathbf{W}_R$ for OTA construction of the control signal $\mathbf{u}_k$, and (ii) $\mathbf{W}_S$ for the sensing task with uplink signal interference.

\subsection{BS Post-processing Matrix for Control Signal Construction}
The aim of dynamical system controlling is to make the system state $\mathbf{x}_k$ approach the designated reference state. We first construct the OTA-combined control objective by integrating the wireless channel coefficients into the widely used quadratic control optimization.  

\subsubsection{OTA-Combined Control Objective}
Without the loss of generality, we consider $\mathbf{x}_k$ as the state error between the system state $\mathbf{x}_k$ and a zero reference \cite{goodwin2001control}. Then, the widely-studied feedback-looped objective function and the corresponding optimization problem is employed here as \cite{lewis2012optimal,goodwin2001control}:
\begin{subequations}
\label{lqr_obj}
\begin{align}
J_k(\mathbf{x}_k)\triangleq\min_{\mathbf{u}_k,\cdots,\mathbf{u}_{k+K-1}}~&\sum_{i=0}^{K-1}\mathbf{x}_{k+i}^H\mathbf{Q}\mathbf{x}_{k+i}+\mathbf{u}_{k+i}^H\mathbf{R}\mathbf{u}_{k+i}\label{quadratic}\\
&\text{s.t.~}(\ref{ode1})\label{ode}
\end{align}
\end{subequations}
where $\mathbf{Q}\in\mathbb{R}_{6N\times 6N}$ and $\mathbf{R}\in\mathbb{R}_{3N\times 3N}$ are predefined positive semi-definite symmetrical matrices.

In linear control cases where the control signal $\mathbf{u}_k$ is linear to the system state $\mathbf{x}_k$ (e.g., the OTA controller from Eq. (\ref{ul-u})), a quadratic form $J_k(\mathbf{x}_k)=\mathbf{x}_k^H\mathbf{P}_k\mathbf{x}_k$ can be founded, with $\mathbf{P}_k$ a symmetrical positive semi-definite matrix \cite{lewis2012optimal}. As such, Eq. (\ref{lqr_obj}) is reformulated as:
\begin{equation}
\begin{aligned}
&J_k(\mathbf{x}_k)=\mathbf{x}_k^H\mathbf{Q}\mathbf{x}_k+\min_{\mathbf{u}_k}\left(\mathbf{u}_k^H\mathbf{R}\mathbf{u}_k+J_{k+1}(\mathbf{A}\mathbf{x}_k+\mathbf{B}\mathbf{u}_k)\right)\\
=&\min_{\mathbf{u}_k}\left(\mathbf{u}_k^H\mathbf{R}\mathbf{u}_k\!+\!(\mathbf{A}\mathbf{x}_k\!+\!\mathbf{B}\mathbf{u}_k)^H\mathbf{P}_{k+1}(\mathbf{A}\mathbf{x}_k\!+\!\mathbf{B}\mathbf{u}_k)\right)\!+\!\text{const},
\end{aligned}
\end{equation}
where $\text{const}=\mathbf{x}_k^H\mathbf{Q}\mathbf{x}_k$ is irrelavant with the optimization variable $\mathbf{u}_k$. In practice, we replace $\mathbf{P}_{k+1}$ with $\mathbf{P}$, the solution of Riccati function that is obtained by iteratively solving $\mathbf{P}\leftarrow\mathbf{Q}+\mathbf{A}^H\mathbf{P}\mathbf{A}-\mathbf{A}^H\mathbf{P}\mathbf{B}(\mathbf{R}+\mathbf{B}^H\mathbf{P}\mathbf{B})^{-1}\mathbf{B}^H\mathbf{P}\mathbf{A}$, with initial $\mathbf{P}=\mathbf{Q}$ \cite{lewis2012optimal,goodwin2001control}.

As we take the OTA design of Eq. (\ref{ul-u}) in the control signal construction, the uplink optimization problem to design post-processing matrix $\mathbf{W}_R$ is formed as:
\begin{subequations}
\label{ul_opt}
\begin{align}
\mathcal{P}_{1}:&\min_{\mathbf{u}_k,~ \mathbf{W}_R}\mathbf{u}_k^T\mathbf{R}\mathbf{u}_k\!+\!\left(\mathbf{A}\mathbf{x}_k\!+\!\mathbf{B}\mathbf{u}_k\right)^H\!\!\mathbf{P}\left(\mathbf{A}\mathbf{x}_k\!+\!\mathbf{B}\mathbf{u}_k\right),\label{lqr2}\\
&\text{s.t.}~(\ref{ul-u})\label{lqr2_c1}.
\end{align}
\end{subequations}

\subsubsection{Closed-Form Control-centric BS Post-processing Matrix}
In this part, we derive a closed-form solution for problem $\mathcal{P}_1$. 
%It is noteworthy that the objective function in Eq. (\ref{lqr2}) involves the accumulated state control error of future predicted system states, i.e., $\mathbf{x}_{k+1}$ to $\mathbf{x}_{K-1}$, which are related to the feedback from the current control signals $\mathbf{u}_k$, e.g., $\mathbf{x}_{k+1}=\mathbf{A}\mathbf{x}_k+\mathbf{B}\mathbf{u}_k$. These feedback loops make the current OTA-FL scheme difficult to be implanted for control objectives, and therefore requires new design to incorporate OTA computation into control theory. the feedback-looped objective function creates a control layer state dependency among different terminals, which differs from the independent or spatio-temporal assumptions typically employed in FL. 
First, using Eq. (\ref{receive1}) and Eq. (\ref{ul-u}), BS constructed control signal $\mathbf{u}_k$ can be further expressed as:
\begin{equation}
\label{cs1}
\mathbf{u}_k=\mathbf{W}_R\cdot\tilde{\mathbf{H}}^{\text{(ul)}}\cdot\mathbf{x}_k+\mathbf{W}_R\cdot\bm{\Xi}\cdot s+\mathbf{W}_R\cdot\bm{\varepsilon}_k, 
\end{equation}
where
$\tilde{\mathbf{H}}^{\text{(ul)}}\triangleq\mathbf{H}^\text{(ul)}\otimes\mathbf{I}_6$ with $\mathbf{H}^\text{(ul)}\triangleq\gamma^\text{(ul)}\left[\mathbf{h}_{1S}^{\text{(ul)}},\mathbf{h}_{2S}^{\text{(ul)}},\cdots,\mathbf{h}_{NS}^{\text{(ul)}}\right]$, and $\bm{\Xi}\triangleq\mathbf{G}\otimes\mathbf{I}_6$.

In Eq. (\ref{cs1}), it is noticeable that the sensing signal $s$ is not related to the UAV control purpose. We hereby make $\mathbf{W}_R\bm{\Xi}\equiv\mathbf{0}$ by representing the range of $\mathbf{W}_R^H$ via the null-space of $\bm{\Xi}^H$. To do this, we pursue compact singular-value decomposition (SVD) of $\bm{\Xi}$ and obtain the compact left singular matrix, denoted as $\bm{\Gamma}$. Then, the orthogonal matrix of $\bm{\Gamma}$, denoted as $\bm{\Gamma}_\bot$ is used to characterize the rows of $\mathbf{W}_R$, i.e., 
\begin{equation}
\label{cs2}
\mathbf{W}_R=\mathbf{W}\cdot\bm{\Gamma}_\bot,
\end{equation}
where $\mathbf{W}$ is the coefficient matrix to represent $\mathbf{W}_R$. By combining Eqs. (\ref{cs1})-(\ref{cs2}), the BS constructed control signal $\mathbf{u}_k$ can be re-written as:
\begin{equation}
\label{cs3}
\mathbf{u}_k=\mathbf{W}\cdot\bm{\Gamma}_\bot\cdot\tilde{\mathbf{H}}^\text{(ul)}\cdot\mathbf{x}_k+\mathbf{W}\cdot\bm{\Gamma}_\bot\cdot\bm{\varepsilon}_k.
\end{equation}

Then, we take the orthogonal design of $\mathbf{u}_k$ from Eqs. (\ref{cs2})-(\ref{cs3}) back into the objective function of Eq. (\ref{lqr2}). To simplify the expression, we denote $\bm{\Upsilon}\triangleq\bm{\Gamma}_\bot\tilde{\mathbf{H}}^\text{(ul)}$. The objective function of Eq. (\ref{lqr2}) is then expressed as:
\begin{equation}
\label{obj5}
\begin{aligned}
&\mathbf{u}_k^H\mathbf{R}\mathbf{u}_k\!+\!\left(\mathbf{A}\mathbf{x}_k\!+\!\mathbf{B}\mathbf{u}_k\right)^H\!\!\mathbf{P}\left(\mathbf{A}\mathbf{x}_k\!+\!\mathbf{B}\mathbf{u}_k\right)\\
=&\mathbf{x}_k^H\bm{\Pi}_{11}\mathbf{x}_k+\mathbf{x}_k^H\bm{\Pi}_{12}\bm{\varepsilon}_k+\bm{\varepsilon}_k^H\bm{\Pi}_{12}^H\mathbf{x}_k+\bm{\varepsilon}_k^H\bm{\Pi}_{22}\bm{\varepsilon}_k,\\
\end{aligned}
\end{equation}
where 
\begin{equation}
\label{obj5_1}
\begin{aligned}
\bm{\Pi}_{11}=&
\bm{\Upsilon}^H\mathbf{W}^H(\mathbf{B}^H\mathbf{P}\mathbf{B}+\mathbf{R})\mathbf{W}\bm{\Upsilon}+\mathbf{A}^H\mathbf{P}\mathbf{B}\mathbf{W}\bm{\Upsilon}\\
&~~~~+\bm{\Upsilon}^H\mathbf{W}^H\mathbf{B}^H\mathbf{P}\mathbf{A}+\mathbf{A}^H\mathbf{P}\mathbf{A},
\end{aligned}
\end{equation}
\begin{equation}
\label{obj5_2}
\begin{aligned}
\bm{\Pi}_{12}=&
\bm{\Upsilon}^H\mathbf{W}^H(\mathbf{B}^H\mathbf{P}\mathbf{B}+\mathbf{R})\mathbf{W}\bm{\Gamma}_\bot+\mathbf{A}^H\mathbf{P}\mathbf{B}\mathbf{W}\bm{\Gamma}_\bot,
\end{aligned}
\end{equation}
and 
\begin{equation}
\label{obj5_3}
\begin{aligned}
\bm{\Pi}_{22}=&
\bm{\Gamma}_\bot^H\mathbf{W}^H(\mathbf{B}^H\mathbf{P}\mathbf{B}+\mathbf{R})\mathbf{W}\bm{\Gamma}_\bot.
\end{aligned}
\end{equation}
As such, the optimization problem $\mathcal{P}_1$ in Eq. (\ref{ul_opt}) is converted to optimize the objective function in Eq. (\ref{obj5}) with respect to the optimization variable $\mathbf{W}$ that constitutes the post-processing matrix $\mathbf{W}_R$, i.e., 
\begin{equation}
\label{lqr2_1}
\mathcal{P}_{1-1}:\min_{\mathbf{W}}~
[\mathbf{x}_k^H~\bm{\varepsilon}^H]
\begin{bmatrix}
\bm{\Pi}_{11} & \sigma\bm{\Pi}_{12}\\
\sigma\bm{\Pi}_{12}^H & \sigma^2\bm{\Pi}_{22}
\end{bmatrix}
\begin{bmatrix}
\mathbf{x}_k \\ \bm{\varepsilon}
\end{bmatrix}
\end{equation}
where $\bm{\varepsilon}$ is the normalized version (with variance $1$) of $\bm{\varepsilon}_k$.

It is noteworthy that in the optimization process of problem $\mathcal{P}_{1-1}$, BS does not know the system state $\mathbf{x}_k$ and the noise vector $\bm{\varepsilon}$. This is because that BS may not be able to recover all UAV states, given the low-rank of channel matrix. To address this, we convert the problem $\mathcal{P}_{1-1}$, and make it irrelevant to the specific UAV state $\mathbf{x}_k$ and noise vector. We do this by minimizing the largest eigenvalue of the matrix that constitutes the quadratic form of $\mathbf{x}_k$, i.e., 
\begin{equation}
\label{lqr4}
\mathcal{P}_{1-2}:~\min_{\mathbf{W}}~\lambda_\text{max}\left(
\begin{bmatrix}
\bm{\Pi}_{11} & \sigma\bm{\Pi}_{12}\\
\sigma\bm{\Pi}_{12}^H & \sigma^2\bm{\Pi}_{22}
\end{bmatrix}
\right)
\end{equation}
In Eq. (\ref{lqr4}), the coefficient matrices can be factorized as:
\begin{equation}
\label{obj5_4}
\begin{bmatrix}
\bm{\Pi}_{11} & \sigma\bm{\Pi}_{12} \\
\sigma\bm{\Pi}_{12}^H & \sigma^2\bm{\Pi}_{22}
\end{bmatrix}
=
\begin{bmatrix}
\bm{\Upsilon}^H\mathbf{W}^H & \mathbf{A}^H\\
\sigma\bm{\Gamma}_\bot^H\mathbf{W}^H & \mathbf{0}
\end{bmatrix}
\cdot
\bm{\Phi}
\cdot
\begin{bmatrix}
\mathbf{W}\bm{\Upsilon} & \sigma\mathbf{W}\bm{\Gamma}_\bot\\
\mathbf{A} & \mathbf{0}
\end{bmatrix}
\end{equation}
with
\begin{equation}
\label{Phi}
\bm{\Phi}\triangleq
\begin{bmatrix}
\mathbf{B}^H\mathbf{P}\mathbf{B}+\mathbf{R} & \mathbf{B}^H\mathbf{P}\\
\mathbf{P}\mathbf{B} & \mathbf{P}
\end{bmatrix}.
\end{equation}
By taking Eqs. (\ref{obj5_4})-(\ref{Phi}) into Eq. (\ref{lqr4}), we convert the optimization problem $\mathcal{P}_{1-2}$ as:
\begin{equation}
\label{lqr3}
\mathcal{P}_{1-3}:\min_{\mathbf{W}}~
\lambda_\text{max}\left(
\begin{bmatrix}
\bm{\Upsilon}^H\mathbf{W}^H & \mathbf{A}^H\\
\sigma\bm{\Gamma}_\bot^H\mathbf{W}^H & \mathbf{0}
\end{bmatrix}
\bm{\Phi}
\begin{bmatrix}
\mathbf{W}\bm{\Upsilon} & \sigma\mathbf{W}\bm{\Gamma}_\bot\\
\mathbf{A} & \mathbf{0}
\end{bmatrix}\right)
\end{equation}
Given that $\bm{\Phi}$ is hermitian and positive semi-definite, there exists $\bm{\Phi}=\bm{\Psi}^H\bm{\Psi}$ via Cholesky factorization. Then, the objective function in $\mathcal{P}_{1-3}$ is converted, given matrix inequalities, as:
\begin{equation}
\label{ineq}
\begin{aligned}
&\lambda_\text{max}\left(
\begin{bmatrix}
\bm{\Upsilon}^H\mathbf{W}^H & \mathbf{A}^H\\
\sigma\bm{\Gamma}_\bot^H\mathbf{W}^H & \mathbf{0}
\end{bmatrix}
\bm{\Phi}
\begin{bmatrix}
\mathbf{W}\bm{\Upsilon} & \sigma\mathbf{W}\bm{\Gamma}_\bot\\
\mathbf{A} & \mathbf{0}
\end{bmatrix}\right)\\
=&\left\|\bm{\Psi}\cdot\begin{bmatrix}
\mathbf{W}\bm{\Upsilon} & \sigma\mathbf{W}\bm{\Gamma}_\bot\\
\mathbf{A} & \mathbf{0}
\end{bmatrix}\right\|_2^2<\left\|\bm{\Psi}\cdot\begin{bmatrix}
\mathbf{W}\bm{\Upsilon} & \sigma\mathbf{W}\bm{\Gamma}_\bot\\
\mathbf{A} & \mathbf{0}
\end{bmatrix}\right\|_F^2\\
\overset{(a)}{=}&\left\|\bm{\Psi}_1\mathbf{W}\bm{\Upsilon}+\bm{\Psi}_2\mathbf{A}\right\|_F^2+\sigma^2\left\|\bm{\Psi}_1\mathbf{W}\bm{\Gamma}_\bot\right\|_F^2\\
\overset{(b)}{=}&\left\|\left(\bm{\Upsilon}^T\otimes\bm{\Psi}_1\right)\text{vec}(\mathbf{W})+\text{vec}(\bm{\Psi}_2\mathbf{A})\right\|_2^2\\
&~~~~~+\sigma^2\left\|\left(\bm{\Gamma}_\bot^T\otimes\bm{\Psi}_1\right)\text{vec}(\mathbf{W})\right\|_2^2.
\end{aligned}
\end{equation}
In Eq. (\ref{ineq}), (a) is by rewriting $\bm{\Psi}=[\bm{\Psi}_1~\bm{\Psi}_2]$, and (b) is due to $\|\cdot\|_F^2=\|\text{vec}(\cdot)\|_2^2$ and Kronecker product computations.

As such, with the help of Eq. (\ref{ineq}), the minimization problem $\mathcal{P}_{1-3}$ in Eq. (\ref{lqr3}) is further transformed to minimize its upper-bound, i.e., 
\begin{equation}
\label{lqr5}
\mathcal{P}_{1-4}:~\min_{\mathbf{W}}~(\ref{ineq}). 
\end{equation}
The closed-form solution of $\mathcal{P}_{1-4}$ can be obtained by letting the derivative of Eq. (\ref{ineq}) to $\text{vec}(\mathbf{W})$ equal $0$, i.e., 
\begin{equation}
\label{W}
\begin{aligned}
\text{vec}(\mathbf{W})=&-\left[\left(\bm{\Upsilon}\bm{\Upsilon}^H+\sigma^2\bm{\Gamma}_\bot\bm{\Gamma}_\bot^H\right)\otimes\left(\bm{\Psi}_1^H\bm{\Psi}_1\right)\right]^{-1}\\
&\cdot\left(\bm{\Upsilon}^T\otimes\bm{\Psi}_1\right)^H\cdot\text{vec}\left(\bm{\Psi}_2\mathbf{A}\right).
\end{aligned}
\end{equation}
With Eq. (\ref{W}), the BS post-processing matrix $\mathbf{W}_R=\mathbf{W}\bm{\Gamma}_{\bot}$ can be computed and assigned for BS. The control signal $\mathbf{u}_k$ is then obtained after the post-processing of BS's received signal $\mathbf{r}_k^{\text{(ul)}}$ according to Eq. (\ref{ul-u}).

\subsubsection{Analysis of Upper-bound Relaxation Gap}
It is noteworthy that the derivation of this closed-form OTA control solution is based on the relaxed upper-bound in $\mathcal{P}_{1-4}$ of original OTA control objective $\mathcal{P}_1$ in Eq. (\ref{ul_opt}). We show that minimizing the relaxed upper-bound in $\mathcal{P}_{1-4}$ does minimize the original OTA control objective. 

It is first seen that $\mathcal{P}_1$ and $\mathcal{P}_{1-1}$ is equivalent, as no relaxation is made from Eq. (\ref{ul_opt}) to Eq. (\ref{lqr2_1}). Then, from $\mathcal{P}_{1-1}$ to $\mathcal{P}_{1-2}$ and its equivalent $\mathcal{P}_{1-3}$, the largest eigenvalue relaxation is used and indeed serves as a reachable upper-bound, for any possible system state $\mathbf{x}_k$ and normalized noise vector $\bm{\varepsilon}$ that may be taken in the minimization problem in $\mathcal{P}_{1-1}$. Next, from $\mathcal{P}_{1-3}$ to the final $\mathcal{P}_{1-4}$, the relaxation in Eq. (\ref{ineq}) is from the norm inequality, i.e., $\bar\omega\cdot\|\cdot\|_F^2<\|\cdot\|_2^2<\|\cdot\|_F^2$ for some $\bar\omega\in(0,1)$, which therefore leads to $\bar\omega\cdot(\ref{ineq})<(\ref{lqr3})<(\ref{ineq})$. As such, by minimizing the objective function Eq. (\ref{ineq}) in $\mathcal{P}_{1-4}$, the optimization problem $\mathcal{P}_{1-3}$ can be minimized and so does the original OTA control problem in $\mathcal{P}_1$.

\subsection{BS Post-processing Matrix for Sensing}\label{uplink-sensing}
The sensing-centric BS post-processing matrix $\mathbf{W}_S$ is to separate echoed sensing signal from the transmitted UAV states. By taking Eq. (\ref{receive1}) into Eq. (\ref{ul-s}), we rewrite the post-processed received sensing signal as:
\begin{equation}
\begin{aligned}
&\mathbf{q}[t]=\mathbf{W}_S\cdot\mathbf{G}\cdot s[t]+\mathbf{W}_S\cdot\sum_{n=1}^N\mathbf{h}_{nS}^{\text{(ul)}}\cdot \gamma^\text{(ul)}x_{k,t}^{(n)}+\mathbf{W}_S\cdot\bm{\varepsilon}_k[t].
\end{aligned}
\end{equation}
We here avoid the interference with the uplink transmitted signal in the second term, by finding the orthogonal (quasi-orthogonal) subspace to the uplink UAV-BS channels, i.e., $\mathbf{W}_S\cdot\left[\mathbf{h}_{1S}^{\text{(ul)}},\mathbf{h}_{2S}^{\text{(ul)}},\cdots,\mathbf{h}_{NS}^{\text{(ul)}}\right]=\mathbf{W}_S\cdot\mathbf{H}^{\text{(ul)}}\approx\mathbf{0}$. 

To achieve this, we first pursue complete SVD of $\mathbf{H}^{\text{(ul)}}$ and find the $d\in\mathbb{N}^+$ left singular vectors that correspond to the $d$ least singular values. We denote matrix $\mathbf{D}$ as the obtained $d$-least left singular vectors. Then, we assign $\mathbf{W}_S=\mathbf{D}^H$, which renders $\mathbf{W}_S\cdot\mathbf{H}^\text{(ul)}=\mathbf{D}^H\cdot\mathbf{H}^\text{(ul)}\approx\mathbf{0}$. Next, the sensing process will be pursued by the MUSIC algorithm, which is detailed in Appendix \ref{Appendix1}.

\subsection{Overall Uplink Algorithm}
\begin{algorithm}[t]
\label{algo1}
\caption{Uplink BS post-processing optimization}
\LinesNumbered
\KwIn {Dynamical system model parameters $\mathbf{A}$ and $\mathbf{B}$, control parameters $\mathbf{Q}$ $\mathbf{R}$ and $\mathbf{P}$, and stacked channel $\mathbf{H}^\text{(ul)}=\gamma^\text{(ul)}[\mathbf{h}_{1S}^{\text{(ul)}},\mathbf{h}_{2S}^{\text{(ul)}},\cdots,\mathbf{h}_{NS}^{\text{(ul)}}]$, $\tilde{\mathbf{H}}^\text{(ul)}=\mathbf{H}^\text{(ul)}\otimes\mathbf{I}_6$ and the stacked echoed channel $\bm{\Xi}=\mathbf{G}\otimes\mathbf{I}_6$. }

Compute $\bm{\Gamma}_{\bot}$ as the orthogonal matrix to the compact left singular matrix of $\bm{\Xi}$\;

Compute $\bm{\Upsilon}=\bm{\Gamma}_{\bot}\cdot\tilde{\mathbf{H}}^\text{(ul)}$ and $\bm{\Phi}$ via Eq. (\ref{Phi})\;

Compute $\bm{\Psi}$ via  Cholesky factorization, i.e., $\bm{\Phi}=\bm{\Psi}^H\bm{\Psi}$\;

Construct $\bm{\Psi}_1$ and $\bm{\Psi}_2$ via $\bm{\Psi}=[\bm{\Psi}_1~\bm{\Psi}_2]$\;

Compute optimal $\mathbf{W}$ via Eq. (\ref{W})\;

Compute the BS post-processing matrix for OTA control signal construction $\mathbf{W}_R$ via Eq. (\ref{cs2})\;

Obtain the BS post-processing matrix for sensing $\mathbf{W}_S$ by the least left singular vectors of $\mathbf{H}^\text{(ul)}$, to make $\mathbf{W}_S\mathbf{H}^\text{(ul)}\approx\mathbf{0}$\; 

\KwOut {BS post-processing matrices $\mathbf{W}_R$ and $\mathbf{W}_S$.}
\end{algorithm}

The overall uplink algorithm for OTA control signal construction and sensing purposes is provided in Algorithm \ref{algo1}. The inputs are the dynamical system model parameters $\mathbf{A}$ and $\mathbf{B}$, the control setting matrices $\mathbf{Q}$ $\mathbf{R}$ and $\mathbf{P}$, the stacked wireless uplink channels $\mathbf{H}^\text{(ul)}$ and $\tilde{\mathbf{H}}^\text{(ul)}$, and the echoed sensing channel $\mathbf{G}$. Step 1 is to separate the received UAV states $\mathbf{x}_k$ and the sensing signal by setting control-centric BS post-processing matrix orthogonal to the sensing channel matrix $\bm{\Xi}$. Steps 2-4 are to compute the intermediate results that will be used to compute the control-centric post-processing matrix. Steps 5-6 are to compute the optimized control-centric BS post-processing matrix $\mathbf{W}_R$ for OTA control signal construction. The control signal $\mathbf{u}_k$ can be constructed over the air at the BS, if the computed control-centric BS post-processing matrix $\mathbf{W}_R$ is deployed, as is shown by Eq. (\ref{ul-u}). Step 7 is to obtain the sensing-centric BS post-processing matrix $\mathbf{W}_S$ for sensing purposes, and the sensing process will be pursued by the MUSIC algorithm, detailed in Appendix \ref{Appendix1}.

\section{Downlink OTA Control Signal Dispatch \& Sensing Optimizations}\label{downlink}
After the uplink OTA construction of control signal $\mathbf{u}_k$ at BS, BS will pursue downlink process to balance the control signal dispatch and its sensing task. In the downlink process, BS aims to (i) send UAVs' their own control signal via OTA mode using the same frequency, and (ii) maintain the object sensing task. From Eqs. (\ref{isac_signal1})-(\ref{receive33}), we are required to optimize the BS precoding matrix $\mathbf{W}_T$, the BS sensing beamforming vector $\mathbf{v}$, and the UAV receiver scaling factor $a$, to balance the transmission and sensing metrics. 

\subsection{Transmission Metrics}
In existing OTA-FL frameworks, all clients receive identical information from the BS via the downlink signal (e.g., global aggregated parameters in OTA-FL). As a result, the transmission metric is typically evaluated using statistical metrics such as receiving SNR or related channel capacity. In contrast, the downlink process of the OTA-control framework requires each UAV to receive a specific portion of the overall control signal, all transmitted by BS over the same frequency.

In this view, we assess the expectation error between UAVs' scaled received signals and the BS transmitted constructed control signal. From Eqs. (\ref{isac_signal1})-(\ref{uav_receive1}), we stack all UAVs' received signals of the each of $3$ successive downlink time-slots as $\mathbf{z}_k[t]\triangleq[z_k^{(1)}[t],z_k^{(2)}[t]\cdots,z_k^{(N)}[t]]^T,~t=1,2,3$, which corresponds to BS transmitted control signals in $3$ successive downlink transmission time-slots, i.e., $\tilde{\mathbf{u}}_k[t]$ in Eq. (\ref{isac_signal1}). The expected transmission error of $t$ downlink transmission slot is therefore expressed as:
\begin{equation}
\begin{aligned}
\label{obj1}
&\mathbb{E}_{\sim s,\bm{\epsilon}_k}\left[\left\|a\cdot\mathbf{z}_k[t]-\tilde{\mathbf{u}}_k[t]\right\|_2^2\right]\\
\overset{(a)}{=}&\left\|\left(a\mathbf{H}^{\text{(dl)}}\mathbf{W}_T-\mathbf{I}_{N_t}\right)\tilde{\mathbf{u}}_k[t]\right\|_2^2+\mathbb{E}_{\sim s}\left(\left\|a\cdot s[t]\cdot\mathbf{H}^{\text{(dl)}}\mathbf{v}\right\|_2^2\!\right)\\
&+\mathbb{E}_{\sim\bm{\epsilon}_k}\left(a\|\bm{\epsilon}_k[t]\|_2^2\right)\\
\overset{(b)}{\leq}&\left\|\left(a\mathbf{H}^{\text{(dl)}}\mathbf{W}_T-\mathbf{I}_{N_t}\right)\right\|_F^2\left\|\tilde{\mathbf{u}}_k[t]\right\|_2^2+a^2\left\|\mathbf{H}^{\text{(dl)}}\mathbf{v}\right\|_2^2+a^2N\sigma^2,
\end{aligned}
\end{equation}
where $\mathbf{H}^\text{(dl)}\triangleq[(\mathbf{h}_{S1}^\text{(dl)})^T,(\mathbf{h}_{S2}^\text{(dl)})^T,\cdots,(\mathbf{h}_{SN}^\text{(dl)})^T]^T$.
In Eq. (\ref{obj1}), (a) is due to that the sensing signal $s[t]$ and the received noise vector $\bm{\epsilon}_k[t]=[\epsilon_k^{(1)}[t],\epsilon_k^{(2)}[t],\cdots,\epsilon_k^{(N)}]^T$ are independent random variables, with both zero expectations, i.e., $\mathbb{E}(s[t])=0$ and $\mathbb{E}(\bm{\epsilon}_k[t])=\mathbf{0}$. (b) is because of the norm inequality, with straightforward computations of the last two terms. 

From Eq. (\ref{obj1}), the expected transmission error of all $3$ transmission time-slots can be expressed as follows:
\begin{equation}
\label{obj4}
\begin{aligned}
&\frac{1}{3}\sum_{t=1,2,3}\mathbb{E}_{\sim s,\bm{\epsilon}_k}\left[\left\|a\cdot\mathbf{z}_k[t]-\tilde{\mathbf{u}}_k[t]\right\|_2^2\right]\\
\overset{(a)}{<}&\left\|\left(a\mathbf{H}^{\text{(dl)}}\mathbf{W}_T-\mathbf{I}_{N_t}\right)\right\|_F^2\frac{\left\|\mathbf{u}_k\right\|_2^2}{3}+a^2\left\|\mathbf{H}^{\text{(dl)}}\mathbf{v}\right\|_2^2+a^2N\sigma^2,
\end{aligned}
\end{equation}
where $(a)$ is due to the fact that the BS transmitted control signal at $3$ time-slots $[\tilde{\mathbf{u}}_k[1],\tilde{\mathbf{u}}_k[2],\tilde{\mathbf{u}}_k[3]]$ is a rearrange of its constructed OTA control signal $\mathbf{u}_k$. The expected transmission error in Eq. (\ref{obj4}) is in terms of the BS precoding matrix $\mathbf{W}_T$, the BS sensing beamforming vector $\mathbf{v}$, and the scaling factor $a$ of all UAV receivers, which can be treated as an objective function for ISAC optimization.

\subsection{Minimum Sensing SNR}
For sensing purposes in the downlink process, we employ the minimum SNR of the received echoed radar signals at BS as the sensing metric. According to Eq. (\ref{receive33}), the minimum sensing SNR corresponding to the $3$ successive BS downlink transmission time-slots are expressed as:
\begin{equation}
\label{snr1}
\begin{aligned}
\text{SNR}^{\text{(dl-echo)}}=&\min_{t=1,2,3}\frac{\mathbb{E}_{\sim s}\left(\|\mathbf{G}\cdot(\mathbf{W}_T\cdot\tilde{\mathbf{u}}_k[t]+\mathbf{v}\cdot s[t])\|_2^2\right)}{\mathbb{E}_{\sim\bm{\varepsilon}_k}\left(\|\bm{\varepsilon}_k[t]\|_2^2\right)}\\
=&\min_{t=1,2,3}\frac{\|\mathbf{G}\mathbf{W}_T\tilde{\mathbf{u}}_k[t]\|_2^2+\|\mathbf{G}\mathbf{v}\|_2^2}{N\cdot\sigma^2},
\end{aligned}
\end{equation}
given the fact that $\mathbb{E}(s[t])=0$. 

To ensure the sensing performance, the sensing SNR should be kept large through the optimization variables $\mathbf{W}_T$ and $\mathbf{v}$, which, however, is conflicting when trying to minimize transmission error in Eq. (\ref{obj4}). In the next part, we will design and solve the corresponding ISAC optimization problem to balance the sensing and transmission performance.

\subsection{ISAC Optimization Problem Formulation}\label{isac_opt}
Combing the transmission error in Eq. (\ref{obj4}) and the sensing SNR in Eq. (\ref{snr1}), the ISAC optimization problem is formulated as follows:
\begin{subequations}
\label{isac1}
\begin{align}
\mathcal{P}_2:&\min_{\mathbf{W}_T,\mathbf{v},a}\frac{1}{3}\sum_{t=1,2,3}\mathbb{E}_{\sim s,\bm{\epsilon}_k}\left[\left\|a\cdot\mathbf{z}_k[t]-\tilde{\mathbf{u}}_k[t]\right\|_2^2\right]\label{isac_obj1}\\
&\text{s.t.}~\text{SNR}^{\text{(dl-echo)}}\geq\gamma_\text{SNR},\label{isac_c1}\\ 
&~~~~\left\|[\mathbf{W}_T]_{m,:}\right\|_2^2+|v_m|^2\leq P_\text{BS},~1\leq m\leq N_t,\label{isac_c2}.
\end{align}
\end{subequations}
where $[\mathbf{W}_T]_{m,:}$ represents the $m$-th row of the matrix $\mathbf{W}_T$, and $v_m$ denotes $m$-th element of the vector $\mathbf{v}$.
In $\mathcal{P}_2$, we minimize the transmission error of Eq. (\ref{isac_obj1}), as well as maintaining the sensing SNR in Eq. (\ref{isac_c1}) larger than a predefined threshold $\gamma_\text{SNR}$. Eq. (\ref{isac_c2}) is to set BS downlink transmission power constrain, i.e., $P_\text{BS}$ for combined precoding matrix and sensing beamforming vector .

\subsection{SCA-based Solution to ISAC Optimization Problem}
To solve the optimization problem $\mathcal{P}_2$, we notice that the objective function in Eq. (\ref{isac_obj1}) is non-convex with respect to the optimization variables $a$, $\mathbf{W}_T$, and $\mathbf{v}$, as seen from Eq. (\ref{obj4}), it contains quadratic forms as $a^2\mathbf{W}_T^H\mathbf{W}_T$ and $a^2\mathbf{v}^H\mathbf{v}$. Then, the constraints from Eq. (\ref{isac_c1}) draws non-convex sets for $\mathbf{W}_T$ and $\mathbf{v}$. 

We first convert the non-convex objective function by combining the optimization variables, i.e., 
\begin{subequations}
\label{com_variables}
\begin{align}
&\tilde{\mathbf{W}}_T\triangleq a\cdot\mathbf{W}_T,\\
&\tilde{\mathbf{v}}\triangleq a\cdot\mathbf{v},\\
&\tilde{a}\triangleq a^2.
\end{align}
\end{subequations}
The original optimization problem $\mathcal{P}_2$ in Eq. (\ref{isac1}) can be thereby rewritten by taking $\tilde{\mathbf{W}}_T$, $\tilde{\mathbf{v}}_T$, and $\tilde{a}$ back into Eqs. (\ref{obj4})-(\ref{snr1}) and Eq. (\ref{isac_c2}), i.e., 
\begin{subequations}
\label{isac1_1}
\begin{align}
\mathcal{P}_{2-1}\!\!:&\min_{\tilde{\mathbf{W}}_T,\tilde{\mathbf{v}},\tilde{a}}\left\|\!\left(\mathbf{H}^{\text{(dl)}}\tilde{\mathbf{W}}_T\!-\!\mathbf{I}_{N_t}\right)\right\|_F^2\!\!\frac{\left\|\mathbf{u}_k\right\|_2^2}{3}\!+\!\left\|\mathbf{H}^{\text{(dl)}}\tilde{\mathbf{v}}\right\|_2^2\!+\!\tilde{a}N\sigma^2\!\label{isac1_1_obj}\\
&\text{s.t.}~\|\mathbf{G}\tilde{\mathbf{W}}_T\tilde{\mathbf{u}}_k[t]\|_2^2+\|\mathbf{G}\tilde{\mathbf{v}}\|_2^2\geq\tilde{a} N\sigma^2\gamma_\text{SNR},~t=1,2,3\label{isac1_1_c1}\\
&~~~~\left\|[\tilde{\mathbf{W}}_T]_{m,:}\right\|_2^2+|\tilde{v}_m|^2\leq \tilde{a}\cdot P_\text{BS},~1\leq m\leq N_t\label{isac1_1_c2}.
\end{align}
\end{subequations}
As such, the non-convexity is only from the constraint in Eq. (\ref{isac1_1_c1}), which draws a non-convex set for optimization variables $\tilde{\mathbf{W}}_T$ and $\tilde{\mathbf{v}}$. 

Next, to address the non-convex optimization problem $\mathcal{P}_{2-1}$ in Eq. (\ref{isac1_1}), we provide a SCA based method. To be specific, we assume to obtain an local optimal $\tilde{\mathbf{W}}_T$ and $\tilde{\mathbf{v}}$ at $i$-th iteration, denoted as $\tilde{\mathbf{W}}_T^{(i)}$ and $\tilde{\mathbf{v}}^{(i)}$. Then, the first-order Taylor expansion based lower-bounds of $\|\mathbf{G}\tilde{\mathbf{W}}_T\tilde{\mathbf{u}}_k[t]\|_2^2$ and $\|\mathbf{G}\tilde{\mathbf{v}}\|_2^2$ can be computed based on the points $\tilde{\mathbf{W}}_T^{(i)}$ and $\tilde{\mathbf{v}}^{(i)}$. The non-convex constrain in Eq. (\ref{isac1_1_c1}) will be further relaxed by making the lower-bound no less than the right-hand side. Noticing that the both $\tilde{\mathbf{W}}_T$ and $\tilde{\mathbf{v}}$ are complex variables, their first-order Taylor expansions are:
\begin{equation}
\label{taylor2}
\begin{aligned}
\left\|\mathbf{G}\tilde{\mathbf{v}}\right\|_2^2\!\geq\!\left\|\mathbf{G}\tilde{\mathbf{v}}^{(i)}\right\|_2^2\!+\!2\Re\left\{\left[\mathbf{G}^T\mathbf{G}^*\left(\tilde{\mathbf{v}}^{(i)}\right)^*\right]^T\!\!\cdot\left(\tilde{\mathbf{v}}-\tilde{\mathbf{v}}^{(i)}\right)\right\}, \\
\end{aligned}
\end{equation}
and 
\begin{equation}
\label{taylor3}
\begin{aligned}
&\left\|\mathbf{G}\tilde{\mathbf{W}}_T\tilde{\mathbf{u}}_k[t]\right\|_2^2\geq\left\|\mathbf{G}\tilde{\mathbf{W}}_T^{(i)}\tilde{\mathbf{u}}_k[t]\right\|_2^2\\
&\!\!+\!2\Re\!\left\{\!\text{vec}^T\!\!\left[\mathbf{G}^T\!\mathbf{G}^*\!\!\left(\!\tilde{\mathbf{W}}_T^{(i)}\!\right)^*\!\!\|\tilde{\mathbf{u}}_k[t]\|_2^2\right]\!\!\cdot\!\!\left[\text{vec}\!\left(\!\tilde{\mathbf{W}}_T\!\right)\!-\!\text{vec}\!\left(\!\tilde{\mathbf{W}}_T^{(i)}\right)\!\right]\!\right\},
\end{aligned}
\end{equation}
where the detailed deduction is provided in Appendix \ref{Taylor_e}. With the help of the linear lower-bounds of the quadratic forms in Eqs. (\ref{taylor2})-(\ref{taylor3}), we relax the non-convex constraint in Eq. (\ref{isac1_1_c1}) of problem $\mathcal{P}_{2-1}$ as:
\begin{equation}
\label{isac3_c1}
\begin{aligned}
&\left\|\mathbf{G}\tilde{\mathbf{v}}^{(i)}\right\|_2^2\!\!+\!\!\left\|\mathbf{G}\tilde{\mathbf{W}}_T^{(i)}\tilde{\mathbf{u}}_k[t]\right\|_2^2\!\!+\!\!2\Re\left\{\!\!\left[\mathbf{G}^T\!\mathbf{G}^*\!\!\left(\tilde{\mathbf{v}}^{(i)}\!\right)^*\right]^T\!\!\!\cdot\!\!\left(\tilde{\mathbf{v}}\!-\!\tilde{\mathbf{v}}^{(i)}\right)\!\!\right\}\\
&\!\!+\!2\Re\!\left\{\!\text{vec}^T\!\!\left[\mathbf{G}^T\!\mathbf{G}^*\!\!\left(\!\tilde{\mathbf{W}}_T^{(i)}\!\right)^*\!\!\|\tilde{\mathbf{u}}_k[t]\|_2^2\right]\!\!\cdot\!\!\left[\text{vec}\!\left(\!\tilde{\mathbf{W}}_T\!\right)\!-\!\text{vec}\!\left(\!\tilde{\mathbf{W}}_T^{(i)}\right)\!\right]\!\right\}\\
\geq&\tilde{a}\cdot N\sigma^2\gamma_\text{SNR}, ~t=1,2,3
\end{aligned}
\end{equation}
which draws affine convex sets to $\tilde{\mathbf{W}}_T$, $\tilde{\mathbf{v}}$, and $\tilde{a}$. 

Leveraging the new convex constraint of Eq. (\ref{isac3_c1}), the SCA based optimization problem is formulated as:
\begin{subequations}
\label{isac3}
\begin{align}
\mathcal{P}_{2-2}:&\min_{\tilde{\mathbf{W}}_T,\tilde{\mathbf{v}},\tilde{a}}(\ref{isac1_1_obj})\\
&\text{s.t.}~(\ref{isac3_c1}),~(\ref{isac1_1_c2}).
\end{align}
\end{subequations}
which is convex and can be solved by CVX Toolbox.

\begin{algorithm}[t]
\label{algo2}
\caption{SCA based optimization of BS downlink precoding matrix $\mathbf{W}_T$, BS sensing beamforming vector $\mathbf{v}$, and UAV receiver scaling factor $a$.}
\LinesNumbered
\KwIn {BS constructed control signal $\mathbf{u}_k$, channel $\mathbf{H}^\text{(dl)}=[(\mathbf{h}_{S1}^\text{(dl)})^T,(\mathbf{h}_{S2}^\text{(dl)})^T,\cdots,(\mathbf{h}_{SN}^\text{(dl)})^T]^T$, echoed sensing channel $\mathbf{G}$, noise variance $\sigma^2$, the minimum BS sensing SNR $\gamma_\text{SNR}$, and BS transmission power $P_\text{BS}$.}

Initialize the optimization variables $\tilde{\mathbf{W}}_T$ and $\tilde{\mathbf{v}}$ at start, i.e., $\tilde{\mathbf{W}}_T^{(0)}$ and $\tilde{\mathbf{v}}^{(0)}$ for $i=0$-th iteration, and assign the stop condition $\iota$\;

\While {$\Delta>\iota$}{
Obtain optimal $\tilde{\mathbf{W}}_T$, $\tilde{\mathbf{v}}$, and $\tilde{a}$ by solving convex problem $\mathcal{P}_{2-2}$ in Eq. (\ref{isac3})\;
Assign $\tilde{\mathbf{W}}_T^{(i+1)}=\tilde{\mathbf{W}}_T$, and $\tilde{\mathbf{v}}^{(i+1)}=\tilde{\mathbf{v}}$\;
$\Delta=\|\tilde{\mathbf{W}}_T^{(i+1)}-\tilde{\mathbf{W}}_T^{(i)}\|_F+\|\tilde{\mathbf{v}}^{(i+1)}-\tilde{\mathbf{v}}^{(i)}\|_2$\;
$i=i+1$\;
}
Obtain $\mathbf{W}_T=\mathbf{W}_T^{(i)}/\sqrt{\tilde{a}}$, $\mathbf{v}=\mathbf{v}^{(i)}/\sqrt{\tilde{a}}$, and $a=\sqrt{\tilde{a}}$\;
\KwOut {Optimized BS downlink precoding matrix $\mathbf{W}_T$, BS sensing beamforming vector $\mathbf{v}$, and actuator receiver scaling factor $a$.}
\end{algorithm}

The overall SCA based algorithm is provided in Algorithm \ref{algo2}. The inputs are the BS constructed control signal $\mathbf{u}_k$ for OTA transmission, the dowinlink channel matrix $\mathbf{H}^\text{(dl)}$, the echoed sensing channel matrix $\mathbf{G}$, the noise variance $\sigma^2$, and the predefined minimum BS sensing SNR $\gamma_\text{SNR}$, and BS transmission power limit $P_\text{BS}$. Step 1 is to initialize the iterative optimization variables $\tilde{\mathbf{W}}_T$ and $\tilde{\mathbf{v}}$ as $\tilde{\mathbf{W}}_T^{(0)}$ and $\tilde{\mathbf{v}}^{(0)}$ at the start $i=0$-th iteration, and assign the stop condition $\iota$. Steps 2-7 are the iterative SCA procedures to obtain the optimal $\tilde{\mathbf{W}}_T^{(i+1)}$, $\tilde{\mathbf{v}}^{(i+1)}$, and $\tilde{a}$ of the problem $\mathcal{P}_{2-2}$. Step 7 is to transform the optimized $\tilde{\mathbf{W}}_T$, $\tilde{\mathbf{v}}$, and $\tilde{a}$ back to the  BS downlink precoding matrix $\mathbf{W}_T$, the BS sensing beamforming vector $\mathbf{v}$, and the UAV receivers' scaling factor $a$, which serve as the outputs. These optimized outputs, adopted in the downlink process detailed in Section \ref{BS-UAV}, ensure that (i) the each UAV can obtain their own control signals for controlling purposes, and (ii) sufficient sensing SNR for continuous radar sensing objective at BS.

\section{Simulations Results}\label{sim}
In this section, we evaluate our proposed ISAC-OTA controller scheme. The performance contains (i) the wireless communication based OTA controller, and (ii) the sensing performance with interference of transmitted OTA signals. The parameter setting is provided in Table \ref{table1}.

\subsection{Metrics Comparison of OTA Controllers}
In this part, we evaluate the control performance in terms of the state controlling errors, with respect to the number of BS antennas, and the setting of minimum sensing SNR.

\subsubsection{OTA control error \& time convergence}
\begin{figure}[!t]
\centering
\includegraphics[width=3.4in]{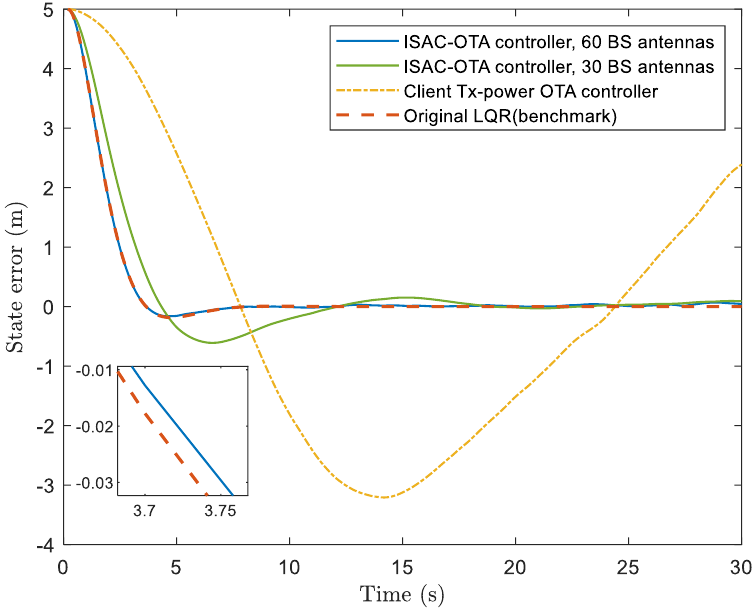}
\caption{Control performance comparison among the proposed ISAC-OTA controller, the existing Tx-power-OTA controller \cite{9641840}, and the original LQR as the benchmark.}
\label{figs1}
\end{figure}

Fig. \ref{figs1} shows the errors of the actual states from the referenced states versus time, provided by our proposed ISAC-OTA controller, the existing UAV Tx-power allocated OTA controller \cite{9641840}, and the classic LQR as benchmark \cite{lewis2012optimal,goodwin2001control} that uses $N$ unique bandwidth to control $N$ UAVs. 
It is first observed that our proposed ISAC-OTA controllers have fewer errors and faster convergence time, as opposed to the existing UAV Tx-power based OTA controller \cite{9641840}. This is because the number of optimization variables for the existing Tx-power-based OTA controller is constrained by the number of each terminal user's Tx antennas, and is insufficient to tune the wireless channels for OTA control purposes. To be specific, in our configuration with $N=50$ UAVs and $6$ states each, Tx-power-based OTA controller can only have $6^2\times50=1800$ precoding variables, which is insufficient to configure the linear control operator of size $(3\times50)\times(6\times50)=45,000$ that maps $\mathbf{x}_k$ to $\mathbf{u}_k$. Compared to the terminal UAV, the BS can be equipped with more antennas for OTA control optimization, suggesting its capability of tuning the wireless channel to better construct and dispatch control signals. 

It is observed that with $N_t=N_r>N$ BS antennas, the control performance of the proposed ISAC-OTA controller is comparable to that of the benchmark LQR’s optimal control. Notably, however, the proposed ISAC-OTA controller achieves this using only a single frequency bandwidth, unlike the benchmark LQR, which requires $N$ unique frequency allocations to control $N$ UAVs.

\begin{table}[!t]
\centering
\caption{Simulation Parameters}
\label{table1}
\begin{tabular}{|l|l|}
\hline
{\bf Parameters} & {\bf Configurations} \\[1mm]
\hline
Number of UAVs & $N=50$ \\
\hline
Control interval \cite{10444964} & $\Delta_t=0.02$s \\
\hline
$\mathbf{Q}$ in control objective & $\mathbf{I}_{N}\otimes\text{diag}([1, 1, 1, 0.1, 0.1, 0.1])$\\
\hline
$\mathbf{R}$ in control objective & $\mathbf{I}_{N}\otimes\text{diag}([1, 1, 1])$\\
\hline
Initial UAV xyz-positions (m)&$x_{1,1}^{(n)},x_{1,2}^{(n)}\in\mathcal{U}(0,100),~x_{1,3}^{(n)}=0$\\
\hline
Number of antennas per UAV& $1$ \\
\hline
UAV Tx power & $0.1$W\\
\hline
BS xyz-position (m)& $(0,0,5)$\\
\hline
Number of BS antennas & $N_t=N_r\in[10,60]$\\
\hline
Uplink scaling factor & $\gamma_\text{ul}=100$\\
\hline
BS Tx power & $P_\text{BS}=1$W\\
\hline
Number of wave paths \cite{9300189} & $L=10$  \\
\hline
Path-loss at $1$m & $C_0=10^{-3}$\\
\hline
LoS exponential loss factor & $\eta_\text{LoS}=2.5$ \\
\hline
NLoS exponential loss factor & $\eta_\text{NLoS}=3$ \\
\hline
Azimuth angles of NLoS paths & $\theta_{nS,l},\theta_{Sn,l}\sim\mathcal{U}[-\pi/2,\pi/2),l>1$  \\
\hline
Sensing object xzy-position (m) & $(10,10,1)$ leads to $\alpha_o=0.7854$ \\
\hline
Echoed sensing gain & $\beta=0.1\cdot C_0 d_{oS}^{-\eta_\text{LoS}}$\\
\hline
Minimum BS sensing SNR & $\gamma_\text{SNR}\in[10,70]$dB\\
\hline
Received noise variance & $\sigma^2=-110$dBW \\
\hline
\end{tabular}
\end{table}

\begin{figure}[!t]
\centering
\includegraphics[width=3.4in]{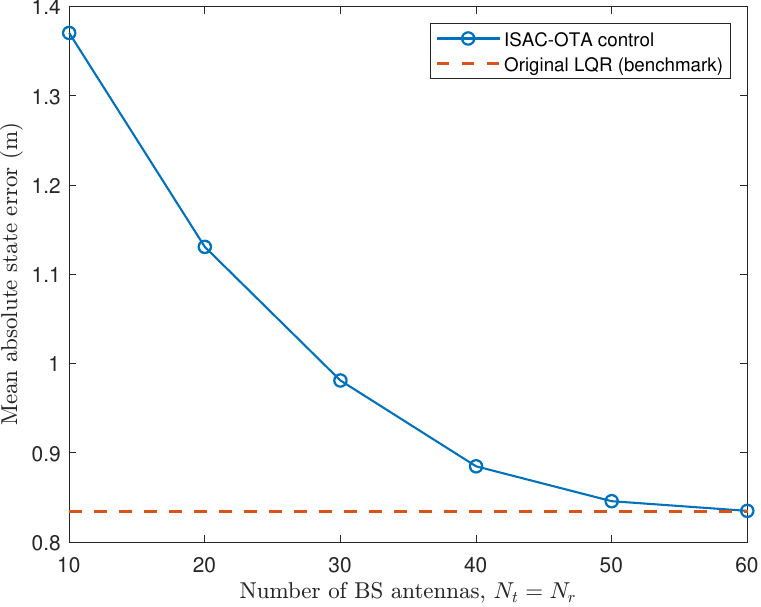}
\caption{Mean absolute state controlling error of proposed ISAC-OTA controller, with respect to different numbers of BS antennas. }
\label{figs2}
\end{figure}

\subsubsection{OTA control error versus number of antennas}
From Fig. \ref{figs1}, we also notice that with the increase of BS antennas (e.g., from $N_t=N_r=30$ to $N_t=N_r=60$), both the errors and the convergence time of our proposed ISAC-OTA controllers decrease. This can be further validated via Fig. \ref{figs2}, which shows a decreased mean absolute state error with respect to the increased number of BS antennas. This is because more BS antennas can provide more uplink and downlink optimization variables to achieve better performances of uplink OTA control signal construction, and downlink control signal dispatch.

\begin{figure}[!t]
\centering
\includegraphics[width=3.4in]{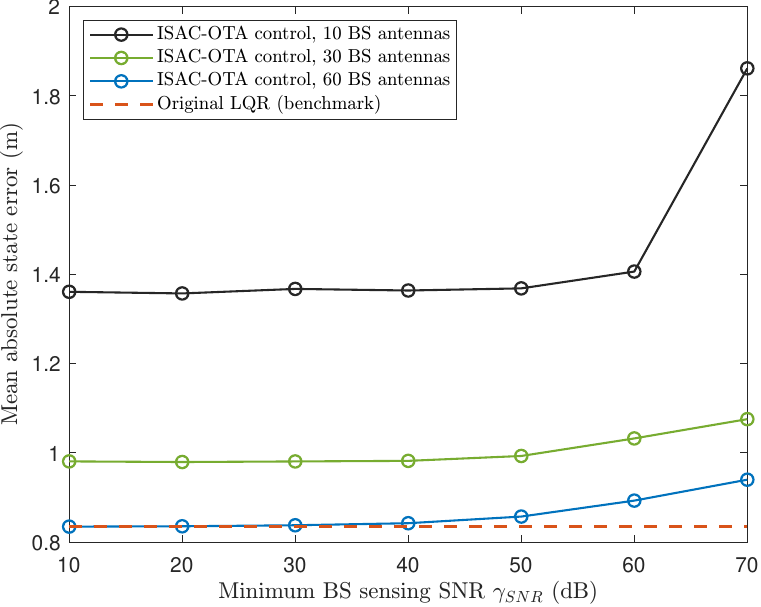}
\caption{Mean absolute state controlling error of proposed ISAC-OTA controller, with respect to different levels of minimum BS sensing SNR $\gamma_\text{SNR}$. }
\label{figs4}
\end{figure}

\subsubsection{OTA control error versus minimum sensing SNR}
We next provide the control performance of the proposed ISAC-OTA controller with respect to the different minimum BS sensing SNR levels set in the downlink process, i.e., $\gamma_\text{SNR}$. It is seen from Fig. \ref{figs4} that as the setting of $\gamma_\text{SNR}$ increases, the mean absolute state errors of the proposed ISAC-OTA controllers go up. This is due to the fact that both the OTA transmission and the sensing task are integrated within the common time, frequency, and spatial resources, rendering the conflict to maintain both of their performances. Yet, with the help of our ISAC optimization design in Eq. (\ref{isac1}) and the SCA-based solution in Algorithm \ref{algo2}, the control accuracy of our proposed ISAC-OTA controller is still remarkable. For instance, when using $60$ BS antennas, the mean absolute state errors are nearly identical to those of the classic LQR algorithm for $\gamma_\text{SNR} \leq 50$ dB. As noted in \cite{10158711}, a minimum sensing SNR of $\gamma_\text{SNR} = 15$ dB is sufficient to guarantee successful sensing. Our proposed ISAC-OTA controller therefore is capable of effectively balancing both sensing and OTA control objectives.

\subsection{BS Sensing Performance}
In this part, we evaluate the BS sensing performance with the interference from the uplink and downlink OTA transmissions. Here, the sensing accuracy is measured by the mean absolute error of the object's angle estimated by the MUSIC algorithm in Appendix \ref{Appendix1}. 

\begin{figure}[!t]
\centering
\includegraphics[width=3.4in]{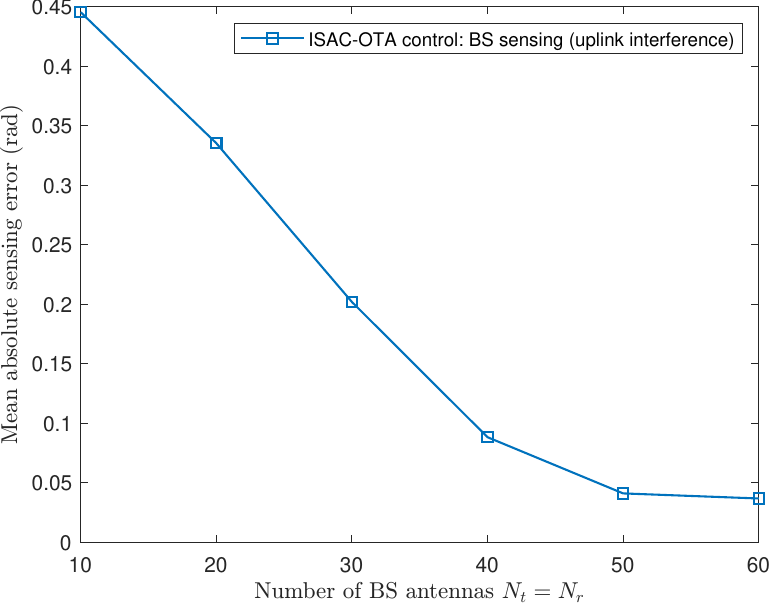}
\caption{Mean absolute sensing error of proposed ISAC-OTA controller in the uplink process, with respect to the number of BS antennas. }
\label{figs3}
\end{figure}

\subsubsection{BS sensing with uplink interference}
Fig. \ref{figs3} provides the sensing performance of the proposed ISAC-OTA scheme, interfered by the uplink OTA transmission signals. It is seen that with the increase of BS antennas (from $10$ to $60$), the sensing error reduces (from $0.45$rad to $0.05$rad). Recall that the design of uplink post-processing matrix $\mathbf{W}_S$ for sensing purposes (in Section \ref{uplink-sensing}) requires to find the orthogonal (quasi-orthogonal) subspace to the uplink UAV-BS channel matrix $\mathbf{H}^\text{(ul)}$ of size $N_r\times N$. Note that the rank of $\mathbf{H}^\text{(ul)}$ may approach to $N$, given the fact that the channels from $N$ UAVs to BS are independent. As such, when the number of BS antennas is less than the number of UAVs, i.e., $N_r<N$, it is less likely to approach $\mathbf{W}_S\mathbf{H}^\text{(ul)}\approx\mathbf{O}$, and therefore leads to more uplink interference to deteriorate the sensing accuracy (e.g., an angle error of $0.2$rad with $N_r=30<50$). Then, when the number of BS antennas is larger than that of UAVs, i.e., $N_r\geq N$, an orthogonal subspace is existed for BS to separate the sensing signal and uplink OTA transmission, which therefore results in small sensing error (e.g., an angle error below $0.05$rad with $N_r=60>N=50$).

\begin{figure}[!t]
\centering
\includegraphics[width=3.4in]{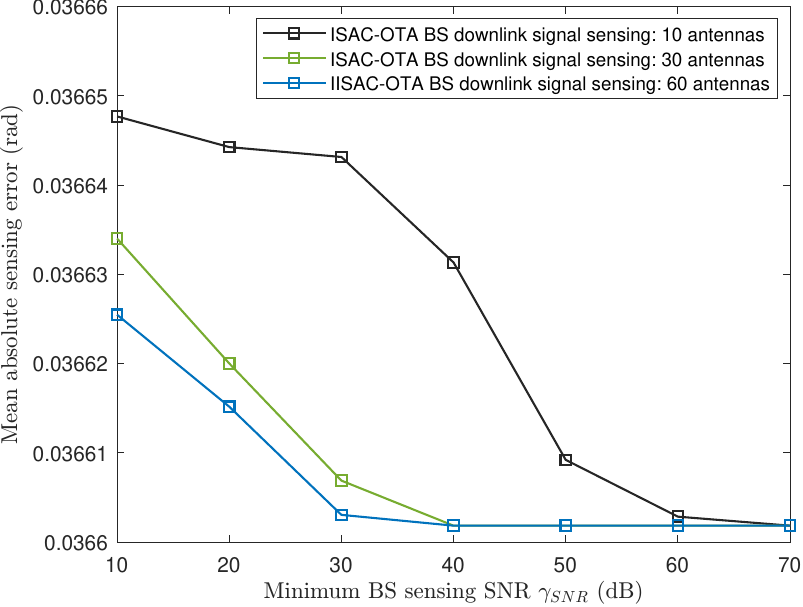}
\caption{Mean absolute sensing error of proposed ISAC-OTA controller in the downlink process, with respect to the different settings of minimum BS sensing SNR $\gamma_\text{SNR}$.}
\label{figs5}
\end{figure}

\subsubsection{BS sensing in downlink process}
Fig. \ref{figs5} illustrates the BS sensing performance in the downlink process. Here, different from the uplink process that the uplink UAV-BS transmission interferes the sensing operation, in the downlink process, BS can use both of its downlink transmission and the continuous radar sensing signal for sensing purposes. In this view, we can first observe an overall lower sensing error as opposed to the sensing operation in the uplink process (e.g., $0.035<0.05$). Then, from Fig. \ref{figs5}, it is also seen that the BS sensing accuracy in the downlink process is affected by both the number of antennas used, and the setting of minimum BS sensing SNR $\gamma_\text{SNR}$. The sensing errors of our proposed ISAC-OTA scheme decrease when larger SNRs or more antennas are allocated. Combining these results, we demonstrate the our proposed ISAC-OTA controller is capable of achieving integrated coordinate UAV swarm control and the sensing tasks within the same frequency bandwidth. 

\begin{figure}[!t]
\centering
\includegraphics[width=3.4in]{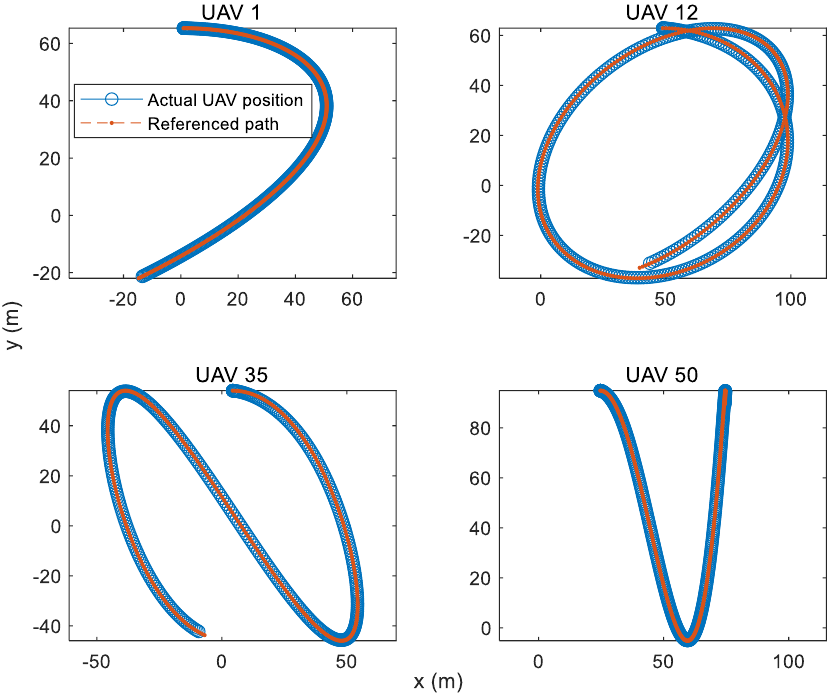}
\caption{Illustration of $4$ (out of $50$) drones referenced paths and actual positions (only xy-plane are shown with constant heights), controlled by our proposed ISAC-OTA controller.}
\label{figs6}
\end{figure}

\subsection{Overall ISAC-OTA Controller Illustration}
We finally show one example of integrated UAV swarm coordinate control and sensing objectives operated by our proposed ISAC-OTA controller. In this example, we use $60$ BS antennas, and the minimum BS sensing SNR is set as $\gamma_\text{SNR}=20$dB. 

In Fig. \ref{figs6}, $4$ out of $N=50$ UAVs are randomly selected for the illustration of the ISAC-OTA controlling performance to control them follow their predefined reference paths. The results show that our proposed ISAC-OTA controller can provide accurate control signals for the swarm of drones to follow the referenced paths.

\begin{figure}[!t]
\centering
\includegraphics[width=3.4in]{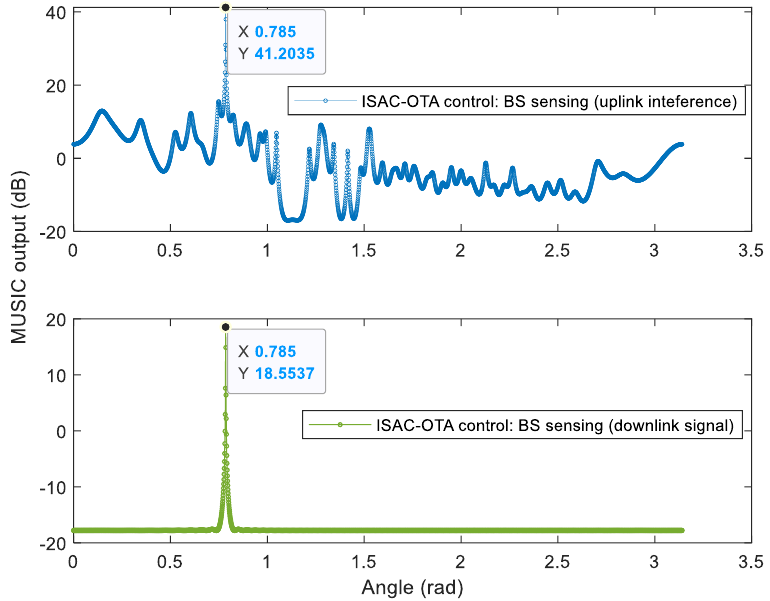}
\caption{Illustration of MUSIC algorithm estimated object angle, with the proposed ISAC-OTA scheme generated input signals: x-coordinate is the angle range in rad, and y-coordinate is the MUSIC output in dB.}
\label{figs7}
\end{figure}

Fig. \ref{figs7} provides an illustration of the MUSIC algorithm (Appendix \ref{Appendix1}) estimated object angle, with the proposed ISAC-OTA scheme generated input signals. Here, x-coordinate represents the angle range from $[0,2\pi]$, and y-coordinate is the MUSIC output provided by the right-hand side of Eq. (\ref{MUSIC output}) in dB. Given that the object of interest is located at $(10, 10, 1)$ relative to the BS at $(0, 0, 5)$, resulting in an azimuth angle from the BS ULA of $\alpha_o = 0.7854$, it is evident from Fig. \ref{figs7} that the proposed ISAC-OTA scheme delivers accurate sensing performance in both the uplink and downlink processes.

\section{Conclusion}\label{conclusion}
OTA controllers leverage the physical superposition of nodes' transmitted signals (states) on the same frequency to generate control signals over the air, avoiding the need of extensive spectrum resources allocated for different nodes. However, existing Tx-power based OTA controllers, operating at each terminal, have two drawbacks. First, the size of the OTA control operator, determined by the product of the transmitted state-space and control signal space dimensions, far exceeds the number of optimization variables in the existing terminal Tx-power-based OTA controller, causing severe control error and slow convergence time. Second, how to integrate the coordinate controller role with the diverse functionalities of BS (e.g., communication, sensing) is not studied.

In this work, the ISAC-OTA controller is proposed for UAV swarm system, where BS is designed to facilitate OTA control signal construction (uplink) and dispatch (downlink), as well as maintaining the object sensing task. In uplink process, two BS post-processing matrices are designed. The control-centric matrix determines a closed-form control signal by solving a feedback-looped control objective that incorporates uplink UAV-BS channels. The sensing-centric matrix spans from the orthogonal subspace of the uplink channels to mitigate transmission interference, ensuring accurate sensing. For the downlink process, we formulate a non-convex optimization problem to minimize transmission error for control signal dispatch while maintaining a minimum sensing SNR. The optimal (or sub-optimal) BS precoding matrix and sensing beamforming vector are determined through the design of the SCA algorithm. The simulation results show (i) comparable control performance of our ISAC-OTA controller with the benchmark LQR control algorithm, without the need to allocate different bandwidths to each UAV, and (ii) high sensing accuracy despite OTA transmission interference. This therefore underscores the potential of utlizing the proposed ISAC-OTA controller for spectrum-efficient cooperative control in future wireless systems.

%\par
%\textbf{Contributions: }
%Z.W., W.H., Y.C. and J.M. developed the idea of the paper. Z.W., Y.B., W.H., C.L., and Y.C developed the OTA control and optimization algorithms. Y.C., W.H., Z.W., and H.S. deduced and checked correctness of equations and availability of algorithms. C.L., Y.C., H.S. and Z.W. provided the modeling, simulation platform and parameters. Y.B., W.H. and J.M. provided guidance on control, communication, sensing metrics and comparative works. J.M., W.H., and Y.C provided guidance on the problem context and impact pathway. All author wrote the paper together. \\

\appendices

\section{MUSIC Sensing Algorithm}\label{Appendix1}
MUSIC algorithm is one super-resolution algorithm for parameter estimation (e.g., angles, time of arrival) of objects of interests, which exploits the orthogonality of signal subspace \cite{sayed2001survey}. For this work, we rewrite BS received signals at successive transmission time-slot as:
\begin{equation}
\bm{\xi}[t]=
\begin{cases}
\mathbf{q}[t],~t=1,\cdots,6 & \text{Uplink post-process in (\ref{ul-s})}\\
\mathbf{r}_k^\text{(dl)}[t],~t=1,2,3 & \text{Downlink process in (\ref{receive33})}
\end{cases}.
\end{equation}
Then, the signal subspace is constructed as:
\begin{equation}
\bm{\Sigma}=\sum_{t}\bm{\xi}\cdot\bm{\xi}^H=\underbrace{\mathbf{F}\cdot\bm{\Lambda}\cdot\mathbf{F}^H}_{\text{signal subspace}}+\underbrace{\mathbf{F}_\bot\cdot\bm{\Lambda}_\bot\cdot\mathbf{F}_\bot^H}_{\text{noise subspace}},
\end{equation}
where $\bm{\Lambda}$ and $\mathbf{F}$ are the $M$ largest eigenvalues and the corresponding eigenvector composed matrix, and $\bm{\Lambda}_\bot$ and $\mathbf{F}_\bot$ are the other rest eigenvalues and the corresponding eigenvector composed matrix. In the case of single object sensing, we have $M=1$. As such, we create the spatial vector spanning from the range of angles of the sensing object, i.e., 
\begin{equation}
\mathbf{a}(\alpha)=
\begin{cases}
\mathbf{W}_S\cdot\bm{\psi}(\alpha, N_r) & \text{Uplink post-process},\\
\bm{\psi}(\alpha, N_r) & \text{Downlink process}.
\end{cases}
\end{equation}
Then the estimated object angle, denoted as $\hat{\alpha}_o$ can be obtained by finding the maximum MUSIC output as follows:
\begin{equation}
\label{MUSIC output}
\hat{\alpha}_o=\argmax_{\alpha\in[0,2\pi)}\frac{1}{\|\mathbf{F}_\bot^H\cdot\mathbf{a}(\alpha)\|_2^2}.
\end{equation}
We evaluate the BS sensing performance during uplink and downlink processes in Simulation section.

\section{Deduction of Complex 1-Order Taylor Expansion}\label{Taylor_e}
We compute the complex first-order Taylor expansion of the scalar function with the complex matrix input in Eq. (\ref{taylor3}). The deduction process of Eq. (\ref{taylor2}) is similar.

To simplify the expression, we abbreviate $\tilde{\mathbf{u}}_k[t]$ at $t$ downlink transmission time-slots of $k$-th control period as $\tilde{\mathbf{u}}$. 
It is seen that the variable of $\tilde{\mathbf{u}}^H\tilde{\mathbf{W}}_T^H\mathbf{G}^H\mathbf{G}\tilde{\mathbf{W}}_T\tilde{\mathbf{u}}$, i.e., $\tilde{\mathbf{W}}_T$, is a complex matrix, rendering the different expression of first-order Taylor expansion, i.e., 
\begin{equation}
\label{taylor11}
\begin{aligned}
&\tilde{\mathbf{u}}^H\tilde{\mathbf{W}}_T^H\mathbf{G}^H\mathbf{G}\tilde{\mathbf{W}}_T\tilde{\mathbf{u}}\geq\tilde{\mathbf{u}}^H\left(\tilde{\mathbf{W}}_T^{(i)}\right)^H\mathbf{G}^H\mathbf{G}\tilde{\mathbf{W}}_T^{(i)}\tilde{\mathbf{u}}\\
&\!\!+\!\!\left(\!\!\frac{\partial\tilde{\mathbf{u}}^H\tilde{\mathbf{W}}_T^H\mathbf{G}^H\mathbf{G}\tilde{\mathbf{W}}_T\tilde{\mathbf{u}}}{\partial\text{vec}(\tilde{\mathbf{W}}_T)}\bigg|_{\tilde{\mathbf{W}}_T^{(i)}}\!\!\right)^T\cdot\left(\!\text{vec}(\tilde{\mathbf{W}}_T)\!-\!\text{vec}\left(\tilde{\mathbf{W}}_T^{(i)}\!\right)\!\right)\\
&\!\!+\!\!\left(\!\!\frac{\partial\tilde{\mathbf{u}}^H\tilde{\mathbf{W}}_T^H\mathbf{G}^H\mathbf{G}\tilde{\mathbf{W}}_T\tilde{\mathbf{u}}}{\partial\text{vec}^*(\tilde{\mathbf{W}}_T)}\bigg|_{\tilde{\mathbf{W}}_T^{(i)}}\!\!\right)^T\cdot\left(\!\text{vec}^*\!(\tilde{\mathbf{W}}_T)\!-\!\text{vec}^*\!\left(\tilde{\mathbf{W}}_T^{(i)}\!\right)\!\right)\\
\end{aligned}
\end{equation}
We then compute the two Wirtinger derivatives of Eq. (\ref{taylor11}). By denoting $f(\tilde{\mathbf{W}}_T)\triangleq\tilde{\mathbf{u}}^H\tilde{\mathbf{W}}_T^H\mathbf{G}^H\mathbf{G}\tilde{\mathbf{W}}_T\tilde{\mathbf{u}}$, the differential of $f(\mathbf{\tilde{\mathbf{W}}_T}):\mathbb{C}^{N_t\times D}\rightarrow\mathbb{R}$ is defined as \cite{hunger2007introduction}:
\begin{equation}
\label{wirtinger111}
\begin{aligned}
&df(\tilde{\mathbf{W}}_T)\\
=&\!\sum_{m,n}\!\!\left(\frac{\partial f(\tilde{\mathbf{W}}_T)}{\partial[\tilde{\mathbf{W}}_T]_{m,n}}\cdot d[\tilde{\mathbf{W}}_T]_{m,n}+\frac{\partial\left(f(\tilde{\mathbf{W}}_T)\right)}{\partial[\tilde{\mathbf{W}}_T]_{m,n}^*}\cdot d[\tilde{\mathbf{W}}_T]_{m,n}^*\!\!\right)\\
=&\left(\frac{\partial f(\tilde{\mathbf{W}}_T)}{\partial\text{vec}(\tilde{\mathbf{W}}_T)}\right)^T \!\!\!d\text{vec}(\tilde{\mathbf{W}}_T)+\left(\frac{\partial f(\tilde{\mathbf{W}}_T)}{\partial\text{vec}^*(\tilde{\mathbf{W}}_T)}\right)^T \!\!\!d\text{vec}^*(\tilde{\mathbf{W}}_T)
\end{aligned}
\end{equation}
On the other hand, the differential can be computed as:
\begin{equation}
\label{wirtinger222}
\begin{aligned}
&df(\tilde{\mathbf{W}}_T)\\
=&\tilde{\mathbf{u}}^H\cdot d\tilde{\mathbf{W}}_T^H\cdot\mathbf{G}^H\mathbf{G}\tilde{\mathbf{W}}_T\tilde{\mathbf{u}}+\tilde{\mathbf{u}}^H\tilde{\mathbf{W}}_T^H\mathbf{G}^H\mathbf{G}\cdot d\tilde{\mathbf{W}}_T\cdot\tilde{\mathbf{u}}\\
=&\text{tr}\left(d\tilde{\mathbf{W}}_T^H\cdot\mathbf{G}^H\mathbf{G}\tilde{\mathbf{W}}_T\tilde{\mathbf{u}}\tilde{\mathbf{u}}^H\right)+\text{tr}\left(\tilde{\mathbf{u}}\tilde{\mathbf{u}}^H\tilde{\mathbf{W}}_T^H\mathbf{G}^H\mathbf{G}\cdot d\tilde{\mathbf{W}}_T\right)\\
=&\text{tr}\!\!\left[\left(\mathbf{G}^H\mathbf{G}\tilde{\mathbf{W}}_T\tilde{\mathbf{u}}\tilde{\mathbf{u}}^H\right)^T\!\!d\tilde{\mathbf{W}}_T^*\!\right]
\!+\!\text{tr}\!\!\left[\left(\mathbf{G}^T\mathbf{G}^*\tilde{\mathbf{W}}_T^*\tilde{\mathbf{u}}^*\tilde{\mathbf{u}}^T\right)^T\!\!d\tilde{\mathbf{W}}_T\!\right]\\
=&\text{vec}^T\left(\mathbf{G}^H\mathbf{G}\tilde{\mathbf{W}}_T\tilde{\mathbf{u}}\tilde{\mathbf{u}}^H\right)\cdot d\text{vec}^*(\tilde{\mathbf{W}}_T)\\
&+\text{vec}^T\left(\mathbf{G}^T\mathbf{G}^*\tilde{\mathbf{W}}_T^*\tilde{\mathbf{u}}^*\tilde{\mathbf{u}}^T\right)\cdot d\text{vec}(\tilde{\mathbf{W}}_T)
\end{aligned}
\end{equation}

As such, by matching Eqs. (\ref{wirtinger111})-(\ref{wirtinger222}), the two Wirtinger derivatives in the first-order Taylor expansion of Eq. (\ref{taylor11}) are derived, i.e., 
\begin{equation}
\label{Wirtinger5}
\frac{\partial\tilde{\mathbf{u}}^H\tilde{\mathbf{W}}_T^H\mathbf{G}^H\mathbf{G}\tilde{\mathbf{W}}_T\tilde{\mathbf{u}}}{\partial\text{vec}(\tilde{\mathbf{W}}_T)}=\text{vec}\left(\mathbf{G}^T\mathbf{G}^*\tilde{\mathbf{W}}_T^*\tilde{\mathbf{u}}^*\tilde{\mathbf{u}}^T\right),
\end{equation}
\begin{equation}
\label{Wirtinger6}
\frac{\partial\tilde{\mathbf{u}}^H\tilde{\mathbf{W}}_T^H\mathbf{G}^H\mathbf{G}\tilde{\mathbf{W}}_T\tilde{\mathbf{u}}}{\partial\text{vec}^*(\tilde{\mathbf{W}}_T)}=\text{vec}\left(\mathbf{G}^H\mathbf{G}\tilde{\mathbf{W}}_T\tilde{\mathbf{u}}\tilde{\mathbf{u}}^H\right).
\end{equation}
In this view, Eq. (\ref{taylor11}) can be further expressed as:
\begin{equation}
\label{taylor22}
\begin{aligned}
&\tilde{\mathbf{u}}^H\tilde{\mathbf{W}}_T^H\mathbf{G}^H\mathbf{G}\tilde{\mathbf{W}}_T\tilde{\mathbf{u}}\geq\tilde{\mathbf{u}}^H\left(\tilde{\mathbf{W}}_T^{(i)}\right)^H\mathbf{G}^H\mathbf{G}\tilde{\mathbf{W}}_T^{(i)}\tilde{\mathbf{u}}\\
+&\text{vec}^T\left(\mathbf{G}^H\mathbf{G}\tilde{\mathbf{W}}_T\tilde{\mathbf{u}}\tilde{\mathbf{u}}^H\right)\cdot\left[\text{vec}^*(\tilde{\mathbf{W}}_T)-\text{vec}^*\left(\tilde{\mathbf{W}}_T^{(i)}\right)\right]\\
+&\text{vec}^T\left(\mathbf{G}^T\mathbf{G}^*\tilde{\mathbf{W}}_T^*\tilde{\mathbf{u}}^*\tilde{\mathbf{u}}^T\right)\cdot\left[\text{vec}(\tilde{\mathbf{W}}_T)-\text{vec}\left(\tilde{\mathbf{W}}_T^{(i)}\right)\right],
\end{aligned}
\end{equation}
which is equivalent to Eq. (\ref{taylor3}) given the fact that the last two terms are conjugate with each other.

\bibliographystyle{IEEEtran}
\bibliography{main.bib}

\end{document}